\documentclass[twocolumn,showpacs,prb,superscriptaddress]{revtex4}
\usepackage{graphicx}
\usepackage{color}\usepackage{enumerate}
\usepackage{amsmath}\usepackage{amssymb}\usepackage{stmaryrd}
\usepackage{multirow}
\usepackage{float}
\usepackage{longtable,booktabs}
\usepackage{subfigure}
\usepackage{ulem}
\usepackage{dsfont}
\usepackage{amssymb}
\usepackage{comment}
\newcommand{\beq}{\begin{eqnarray}}
\newcommand{\eeq}{\end{eqnarray}}
\usepackage{bm}
\usepackage{braket}
\usepackage[pdftex,colorlinks=true,linkcolor=blue,citecolor=blue]{hyperref}
\usepackage{soul}
\usepackage{tikz}

\def\swne{
    \tikz[baseline=0.4ex]{
        \draw (0,0) -- (0,2ex) -- (2ex,2ex) -- (2ex,0) -- (0,0);
        \fill (0,0) circle (2pt);
        \fill (2ex,2ex) circle (2pt);
        \draw[fill=white] (0,2ex) circle (2pt);
        \draw[fill=white] (2ex,0) circle (2pt);
    }
}
\def\nwse{
    \tikz[baseline=0.4ex]{
        \draw (0,0) -- (0,2ex) -- (2ex,2ex) -- (2ex,0) -- (0,0);
        \fill (0,2ex) circle (2pt);
        \fill (2ex,0) circle (2pt);
        \draw[fill=white] (2ex,2ex) circle (2pt);
        \draw[fill=white] (0,0) circle (2pt);
    }
}

\begin{document}
\title{Theory of Dipole Insulators}

\author{Oleg Dubinkin}
\email[]{olegd2@illinois.edu}
\author{Julian May-Mann}
\email[]{maymann2@illinois.edu}
\author{Taylor L. Hughes}
\email[]{hughest@illinois.edu}
\affiliation{Department of Physics and Institute for Condensed Matter Theory, University of Illinois at Urbana-Champaign, IL 61801, USA}
\begin{abstract}
Insulating systems are characterized by their insensitivity to twisted boundary conditions as quantified by the charge stiffness and charge localization length. The latter quantity was shown to be related to the expectation value of the many-body position operator and serves as a universal criterion to distinguish between metals and insulators. In this work we extend these concepts to a new class of quantum systems having conserved charge and dipole moments. We refine the concept of a charge insulator by introducing notions of multipolar insulators, e.g., a charge insulator could be a dipole insulator or dipole metal. We develop a universal criterion to distinguish between these phases by extending the concept of charge stiffness and localization to analogous versions for multipole moments, but with our focus on dipoles. We are able relate the dipole localization scale to the expectation value of a recently introduced many-body quadrupole operator. This refined structure allows for the identification of phase transitions where charge remains localized but, e.g., dipoles delocalize. We illustrate the proposed criterion using several exactly solvable models that exemplify these concepts, and discuss a possible realization in cold-atom systems.

\end{abstract}

%\pacs{}
\maketitle\normalem
\section{Introduction}
Kohn's seminal work on the theory of insulators pinpointed electron localization as the origin of insulating states\cite{kohn1964}. In this work he proposed that a fundamental characteristic of insulators is their insensitivity to twisted boundary conditions, and he quantified this through the  charge stiffness (Drude weight)
\begin{equation}
D_{c}=\frac{1}{V}\frac{\partial^2 E_0(k)}{\partial k^2},
\end{equation}\noindent where $E_0(k)$ is the ground state energy as a function of $k$, the twisted boundary condition phase, and $V$ is the volume. Kohn's criterion for distinguishing insulators from metals is that $D_c\to 0$ in the thermodynamic limit for insulators, and is non-vanishing for metals. Decades later, connections were made between the charge stiffness and superfluid stiffness leading to criteria that distinguish metals, insulators, and superconductors\cite{scalapino1993}. 

More recently Kohn's criterion was re-addressed in the works of Refs. \onlinecite{resta1999,aligia1999,souza2000} which augmented Kohn's charge stiffness with the charge localization length $\xi_c.$ These latter references also reaffirmed the role of localization in distinguishing insulating states from metallic states. Remarkably, the localization length $\xi_c$ is related to the magnitude of the ground-state expectation value of the many-body operator\cite{resta1998}
\begin{equation}
U_{X}=\exp\left[\frac{2\pi i\hat{X}}{L_x}\right],
\end{equation}\noindent where $\hat{X}=\sum_{a=1}^{N} \hat{x}_a$ is the many-body position operator. Given a ground state $\vert \Psi_0\rangle,$ the quantity $z_{X}=\langle \Psi_0\vert U_{X}\vert \Psi_0\rangle$ can be used to determine the localization length in the $x$-direction from its magnitude $\vert z_X\vert$\cite{resta1999,aligia1999}, while its complex phase is proportional to the $x$-component of the charge polarization\cite{resta1998}. The works of Refs. \onlinecite{kudinov1999,souza2000} went on to relate $\vert z_X\vert$, and hence $\xi_c,$ to the fluctuations of the $x$-component of the polarization, which, through the fluctuation-dissipation theorem can be tied to the conductivity. This collection of  groundbreaking work set up a universal criterion for the distinction between metals and insulators from the many-body expectation value: $\vert z_X\vert\to 1$ (0) for insulators (metals) in the thermodynamic limit. The physical interpretation of this quantity is made by connecting $\vert z_X\vert$ to $\xi_c$ (which tends to a finite value for insulators and infinity for metals). Furthermore, from fluctuation-dissipation, the dipole fluctuations (which tend to $\xi_c^2$ in an insulator and infinity in a metal) are tied to the DC conductivity (which tends to zero for insulators, or non-zero in metals) in the thermodynamic limit\cite{resta1999,aligia1999,souza2000}.

In this article we extend these concepts to a new class of quantum systems that more naturally supports dipole transport instead of charge transport. Hence, our focus is on systems with both conserved charge \emph{and} conserved dipole moments. Given a system that is a \emph{charge} insulator, we develop criteria for distinguishing classes of matter based on whether they are dipole conductors or dipole insulators using suitably modified versions of the concepts mentioned above. The motivation for this work is based on recent developments in some classes of matter that are more aptly described in terms of dipole dynamics rather than charge dynamics. This includes some fracton phases of matter\cite{Haah2011-ny,Halasz2017-ov,Vijay2016-dr,Vijay2015-jj,Chamon2005-fc,shirley2018fractional,Slagle2017-ne,Ma2017-qq,Hsieh2017-sc,Vijay2017-ey,Slagle2017-gk,Williamson2016-lv,Ma2017-cb}, and some multipole band insulators\cite{benalcazar2016,benalcazar2017}. Our primary focus will be on fracton matter, which is characterized by the lack of mobility of the fundamental charges in the system and the  constrained dynamics of multipole objects\cite{Ma2017-qq,Slagle2017-gk,Ma2017-cb,pretko2017fracton,ma2018fracton,bulmash2018higgs,Slagle2017-la,shirley2017fracton,prem2018pinch,slagle2018symmetric,gromov2017fractional,pai2018fractonic,pretko2017subdimensional,ma2018higher,pretko2017generalized,pretko2017finite,you2018symmetric,you2018subsystem,you2018symmetric,devakul2018strong,devakul2018fractal}. Hence, fractons are inherently charge insulators, but since some fracton phases have locally conserved dipole moments\cite{pretko2017subdimensional}, it is natural to ask if we can distinguish these charge insulator phases based on whether they are conductors or insulators of \emph{dipoles}. Instead of taking a gauge theory approach to address these questions (see Ref. \onlinecite{ybh2019} for some related discussion along these lines), we will take the technology discussed above for charge insulators and adapt it to describe higher multipole systems. 

Our article is organized as follows. First we provide definitions/criteria to distinguish dipole metals and insulators by defining a dipole stiffness $D_d$ and dipole localization scale $\lambda_d$. We show how $\lambda_d$ is related to the ground-state expectation value of a recently proposed many-body twist operator\cite{wheeler2018many,gil2018}
\begin{equation}
U_{XY}=\exp\left[\frac{2\pi i \widehat{XY}}{L_x L_y}\right],\;\; \widehat{XY}=\sum_{a}^{N} \hat{x}_a\hat{y}_a,
\end{equation} and subsequently to the fluctuations of the quadrupole moment and dipole conductivity. Additionally, we show that the Berry phase associated with the twist implemented by $U_{XY}$ can be used to distinguish between the dipole insulating phases with different overall quadrupolar polarizations. We provide some exactly solvable models on which we can test our criteria, and then discuss a possible physical realization for a dipole metal/insulator in cold-atom systems with ring-exchange interactions. We point out that such systems allow phase transitions where the charge gap remains open, but the dipole gap closes (alternatively where the dipole fluctuations remain finite, but the quadrupole fluctuations diverge). Finally, we conclude with a short discussion on the possible application of our work to higher multipole band insulators and dipole superconductors.

\section{Kohn analogy}
The theory of charge transport relies on the global conservation of charge which manifests in a symmetry transformation of charged operators $e^{iq\alpha}$  where $\alpha$ is a constant and $q$ is the charge. Similarly, if we want to discuss dipole transport we need both global charge and dipole conservation, i.e., both the global particle number and dipole moment are fixed. The latter condition manifests in an invariance under symmetry transformations of the form $e^{i\boldsymbol{\alpha}\cdot \bf{x}}$, where $\boldsymbol{\alpha}$ is a constant vector, i.e., these transformations are exponentials of linear functions of the position coordinates.\cite{griffin2015,pretko2017subdimensional,gromov2019} 
A generic Hamiltonian with dynamics that obeys these conservation laws must commute with both types of transformations. 

To determine the conducting/insulating properties of such systems we need to consider transport in response to applied fields. Hence, we need to couple our system to a background gauge field. From conventional electromagnetism we expect dipole charges and currents to (minimally) couple to the \emph{derivative} of the gauge field $\partial_i A_j$ via $d_i \partial_i A_j$, where $d_i$ is the dipole moment vector.  Alternatively, in the study of fractonic phases of matter it has been shown that it is often natural to couple dipolar excitations to symmetric, rank-2 gauge fields\cite{pretko2017subdimensional,pretko2018gauge}. The rank-2 gauge fields $A_{ij}$ obey a gauge transformation $A_{ij} \rightarrow A_{ij} + \partial_i \partial_j \gamma$, where $\gamma$ is an arbitrary function. 

For this article we will focus on the rank-2 coupling as the models we consider naturally couple to a rank-2 gauge field. The type of gauge coupling one should consider, i.e., rank-1 or rank-2, is context dependent. On one hand, if the microscopic composition of a dipole into constituent particles can be probed, then coupling to a rank-1 vector potential may be more appropriate, since any processes involving charged particle dynamics will violate the rank-2 gauge invariance. On the other hand, in situations where charged particle dynamics are completely frozen, such as in fracton models and the models we consider below, then rank-2 gauge fields may be more natural.  We provide more discussion of this issue, and mention some subtle distinctions between rank-1 and rank-2 couplings, in Appendix \ref{app:rank1rank2}.

%The derivative of a gauge field $\partial_i A_j$ has the gauge transformation $\partial_i A_j \rightarrow \partial_i A_j + \partial_i \partial_j \gamma$, where $\gamma$ is an arbitrary function.

%\textbf{JMM: I don;t see why we need to include the Pretoko Kintetic term. All we need is how the $\Phi$ and $A_{ij}$ transform. OSD: it certainly helps us to provide an explicit argument that Uxy acts on the Hamiltonian the way we claim it does.} 
While most of the explicit work that couples rank-2 gauge fields to matter has been focused on discrete lattice models, Ref. \onlinecite{pretko2018gauge} determined a minimal coupling prescription for a gauge-covariant kinetic operator in a continuum theory \begin{equation}
    D_{ij}[\Phi]=\Phi\partial_i \partial_j \Phi - \partial_i\Phi\partial_j\Phi - iA_{ij}\Phi^2.
    \label{eqn:kin_op}
\end{equation} This operator acts on a charged matter field $\Phi,$ and includes coupling to a rank-2 symmetric gauge field $A_{ij},$ where a rank-2 gauge transformation acts as
\begin{equation}
    \begin{split}
    &\Phi\to \text{e}^{i\gamma(\textbf{x})}\Phi,\;\;\; D_{ij}[\Phi]\to e^{2i\gamma (\textbf{x})}D_{ij}[\Phi]\\
    & A_{ij}(\textbf{x})\to A_{ij}(\textbf{x})+\partial_i \partial_j \gamma(\textbf{x}).
    \end{split}
    \label{}
\end{equation}

Let us now focus on a class of 2D Hamiltonians built from only the $D_{xy}$ kinetic operators and potentials that depend on $\Phi.$ It is straightforward to generalize our considerations to include other $D_{ij}$ components and to treat other dimensionalities.
We can start with a many-body, dipole-conserving Hamiltonian $\mathcal{H}^d$  with the ground state $\Psi_0(x_1,y_1;x_2,y_2...)$. Importantly, since $\mathcal{H}^d$ commutes with dipole transformations of the form $e^{i\alpha_j\hat{X}^j},$ then it also commutes with the twist operators $U_X$ and $U_Y.$ Thus, the eigenstates of $\mathcal{H}^d$ can be chosen to simultaneously diagonalize $U_X$ and $U_Y$ since they commute with each other. A physical consequence, drawn from Ref. \onlinecite{souza2000}, is that these eigenstates have vanishing dipole fluctuations, and hence vanishing \emph{charge} localization length. Explicitly, we have for $U_X$:
\begin{equation}
U_X\Psi_0=\exp\left[\frac{2\pi i\hat{X}}{L_x}\right]\Psi_0=\exp(2\pi i p^x)\Psi_0,
\end{equation}
where $p^x/L_y$ is the polarization in the $x$-direction (where $p_x$ is defined modulo $1$), and similarly for $U_Y$ and the corresponding $p^y.$ 

In analogy with Kohn's work\cite{kohn1964}, let us consider the effects of shifting the rank-2 gauge potential in the Hamiltonian by a constant $A_{xy}\to A_{xy}+\mathfrak{q}.$ The field $A_{xy}$ minimally couples to a dipole current density $j_{xy}\equiv j_d$ that captures the flow of $x$-dipole in the $y$ direction, and $y$-dipole in the $x$-direction. Thus, in analogy to the charge current we can write
\begin{equation}
     j_d(\mathfrak{q})= -\frac{1}{V}\frac{\partial E(\mathfrak{q})}{\partial \mathfrak{q}},
    \label{eq:dipolecurrent}
\end{equation} where $E(\mathfrak{q})$ is the ground state energy of $\mathcal{H}^d$ with $A_{xy}$ shifted by $\mathfrak{q}.$ Using linear response, we can also formulate a dipole conductivity as 
\beq
j_d = \sigma_d E_{xy},
\eeq
where $E_{xy} = \partial_t A_{xy}$ is the rank-2 analog of the electric field. 
In Appendix \ref{app:linearresponse} we show that we can define a dipole stiffness that is directly related to this dipole conductivity as:
\begin{equation}
    \pi\lim_{\omega \rightarrow 0}\text{Im}\omega\sigma_d(\omega) \equiv D_d,
    \label{}
\end{equation}
where 
\begin{equation}
D_d= - \frac{\pi}{V}\left.\frac{\partial^2 E(\mathfrak{q})}{\partial \mathfrak{q}^2}\right|_{\mathfrak{q}=0}.
    \label{eq:dipolestiffness}
\end{equation}
Completing the analogy with charge currents, we propose to use this quantity to distinguish between \emph{dipole} metals and insulators: for a regular charge insulator we expect that $D_d$ either vanishes, in the case of a dipole insulator, or takes a finite value in the case of a dipole metal.
We note that we have only coupled the system to the $A_{xy}$ gauge-field component, and in general, there will be analogous quantities for the other components of $A_{ij}$.

%\textbf{JMM: This statement is not correct OSD: I agree. Nonetheless, the polarization is constant and I can fix it to be integer by choosing appropriate positions for background ions. So the logic holds. I’ll change the wording of the following sentence. JMM: But that means it can always be fixed to be 0, and the rest of the results are trivial OSD: Not exactly. The polarization is defined mod 1, so the result must hold for 0, 1, 2 etc. That’s why I emphasized that p is integer.} 
%Note that the change in polarization is the physically meaningful quantity, not the polarization itself. For a system that conserves dipole moment we can fix polarizations $p^x$ and $p^y$ to be integer-valued by, for example, picking an appropriate distribution of the background positive charge of ions.    
%This phase factor can be removed via an appropriate phase rotation $e^{i\vec{\alpha}\cdot \vec{x}}$ allowed by the dipole conservation condition present in the system and thus we can fix the polarizations in both $x$ and $y$ directions to take integer values.

To lead into the next section, let us now make the connection between the rank-2 gauge field shift and the many-body twist operator $U_{XY}(\mathfrak{q})=e^{-i\mathfrak{q} \widehat{XY}}.$ If we act on the class of Hamiltonians we are considering, we can define  
\begin{equation}
    \mathcal{H}^d(\mathfrak{q})=U_{XY}^{-1}(\mathfrak{q})\mathcal{H}^d U_{XY}(\mathfrak{q}),
\end{equation} where $\mathcal{H}^d(\mathfrak{q})$ differs from $\mathcal{H}^d$ by the replacement $A_{xy}\to A_{xy}+\mathfrak{q}$ in the kinetic terms (\ref{eqn:kin_op}). Indeed the $U_{XY}(\mathfrak{q})$ operator obeys $[D_{ij},U_{XY}(\mathfrak{q})]\Phi=-i\mathfrak{q}\Phi^2$ and thus acts to shift $A_{xy}$ in Eq. \ref{eqn:kin_op}.\footnote{A connection between the action of $U_{XY}(\mathfrak{q})$ and related boundary condition twists, analogous to the usual twisted boundary conditions that are generated by, for example, $U_{X}$ for charge insulators, is a non-trivial open question that we leave to future work.}

We find another interesting application of the twisted Hamiltonian  by starting with a Hamiltonian $\mathcal{H}^d$ that has a charge-neutral, unpolarized ground state $\Psi_0$ that is a charge and dipole insulator. Then one can consider the ``instantaneous" eigenstates of $\mathcal{H}^d(\mathfrak{q})$ that are given by $\ket{\Psi_{\mathfrak{q}}}=U_{XY}^{-1}(\mathfrak{q})\ket{\Psi_0}$. Treating $\mathfrak{q}$ as a small, slowly varying parameter, we can write a perturbative expansion for these states as:
\begin{equation}
%\begin{split}
    \ket{\Psi_{\mathfrak{q}}}=U^{-1}_{XY}(\mathfrak{q})\ket{\Psi_0}\approx \text{e}^{i\gamma_{\mathcal{Q}}(\mathfrak{q})}\ket{\Psi_0}+...
    %-\mathfrak{q}\hbar\sum_{j\neq 0}\ket{\Psi_j}\frac{\bra{\Psi_j}\hat{J}_d\ket{\Psi_0}}{E_0-E_j}\right)
    %\end{split}
\end{equation}
where we kept the phase factor $\gamma_{\mathcal{Q}}(\mathfrak{q})$ that is fixed in the initial state to vanish: $\gamma_{0}(\mathfrak{q})=0$. Then it is natural to introduce a Berry phase in the one-parameter space spanned by $\mathfrak{q}$:
\begin{equation}
    \gamma_{\mathcal{Q}}=\text{Im}\int_{0}^{2\pi/L_x L_y} d\mathfrak{q}\langle\Psi_{\mathfrak{q}}|\partial_{\mathfrak{q}}|\Psi_{\mathfrak{q}}\rangle.
\end{equation} 
where $\gamma_{\mathcal{Q}}\equiv\gamma_{\mathcal{Q}}(2\pi/L_x L_y)$. 

Now let us provide a physical interpretation of this quantity. We note that twisting process can be thought of as an adiabatic evolution of our system from one with $A_{xy}=0$ to a system with $A_{xy}=2\pi/L_x L_y$. We can carry out this process via a time-dependent rank-2 gauge field over a large period of time $T$:
\begin{equation}
    A_{xy}(t)=\frac{2\pi}{L_x L_y}\frac{t}{T}.
\end{equation}
In other words, we turn on a constant rank-2 electric field $E_{xy}=-\partial_t A_{xy}=-2\pi/L_x L_y T$ and track the ground state evolution over a time period $T$. On the other hand, away from the boundaries of the system, $A_{xy}$ is equivalent to the spatial gradient of a regular vector-potential $A_i$ and so, the $E_{xy}$ field has a natural interpretation in terms of the ordinary rank-1 electric fields:
\begin{equation}
\begin{split}
    E_{xy}=\frac1{2\hbar}\left(\partial_x E_y+\partial_y E_x\right).
\end{split}
\end{equation}
Assuming that the ground state of our system is one of a charge-neutral, unpolarized insulating system, the gradient of an electric field couples to the  quadrupole moment resulting into a phase factor:
\begin{equation}
\begin{split}
    \gamma_{\mathcal{Q}}&=\frac{1}{\hbar}\int_0^T dt\ Q_{xy}\frac12(\partial_xE_y+\partial_yE_x)\\
    &=\int_0^T dt\ Q_{xy} E_{xy}=\int_0^T dt\ Q_{xy}\partial_t A_{xy}(t)=\frac{2\pi Q_{xy}}{L_x L_y},
\end{split}
\end{equation}
where $Q_{xy}$ is the  $xy$ quadrupole moment which we assumed to be static.
Thus, the rank-2 Berry phase $\gamma_{\mathcal{Q}}$ naturally corresponds to the quadrupolar polarization, which might have been anticipated from the results of Refs. \onlinecite{wheeler2018many}, \onlinecite{gil2018} where the twist operator $U_{XY}$ was first introduced.

\section{Dipole Localization and Quadrupole Fluctuations}
Coming back to our discussion of dipole metals and dipole insulators, we will now show how the twist operator $U_{XY}=U_{XY}(\mathfrak{q}=2\pi/L_x L_y)$ can be used to determine a dipole localization scale $\lambda_d$ (with units of area) in a many body system. The properties of $\lambda_d$ in the thermodynamic limit also lead to a criterion to distinguish dipole metals from insulators, solely in terms of the ground state localization properties. 

In order to define $\lambda_d$ we still enforce dipole conservation so that the ground state $\vert \Psi_0\rangle$ of $\mathcal{H}^d$ is an eigenstate of the $U_X$ and $U_Y$ operators. Now let us consider the expectation value $z_{XY}=\langle \Psi_0\vert U_{XY}\vert\Psi_0\rangle\equiv \langle U_{XY}\rangle_0$ for the \emph{single-dipole} state $\ket{\Psi_0}$. Assuming that the dipole is localized on a scale much smaller than the system size $L_x L_y,$ we can expand the expectation value in the thermodynamic limit
\begin{eqnarray}
    z_{XY}&=&1+\frac{2\pi i\langle \widehat{XY}\rangle_0}{L_x L_y}-\frac{4\pi^2 \langle \widehat{XY}  \widehat{XY}\rangle_0}{L_x^2 L_y^2}\nonumber\\&+& O(1/L_x^3 L_y^3).
\end{eqnarray} From this expression we can read off that 
\begin{eqnarray}
\frac{1}{2\pi}{\rm{Im}}\log z_{XY}&\approx&\frac{\langle \widehat{XY}\rangle_0}{L_x L_y}=q_{xy},\\
\log\vert z_{XY}\vert^2&\approx&-\frac{4\pi^2}{L_x^2 L_y^2}\left[\langle \widehat{XY}\widehat{XY}\rangle_0-\langle \widehat{XY}\rangle_{0}^2\right],\label{eq:quadfluct}
\end{eqnarray}\noindent where the approximation becomes exact in the thermodynamic limit. These results indicate that the phase of $z_{XY}$ is the quadrupole density $q_{xy}$ (as expected from Refs. \onlinecite{wheeler2018many,gil2018}), and that we can define a `dipole correlation area'
\begin{equation}
    \lambda_{d}^2\equiv -\frac{L_x^2 L_y^2}{4\pi^2}\log\vert z_{XY}\vert^2\approx \langle \widehat{XY}\widehat{XY}\rangle_0-\langle \widehat{XY}\rangle_{0}^2,
    \label{eq:DipoleCorA}
\end{equation} where $\lambda_d$ has units of \emph{area}. From the right hand side we see that $\lambda_d$ is capturing the fluctuations of $q_{xy},$ e.g., the spread of $x$-oriented dipole in the $y$-direction and vice-versa. We remind the reader that since the ground state is an eigenstate of $U_X, U_Y,$ the \emph{dipole} fluctuations $\langle \widehat{X}^2\rangle_0=\langle \widehat{Y}^2\rangle_0=0.$

Following Resta\cite{resta1999} we propose a following extension of this quantity for the state $\ket{\Psi_{N_d}}$ containing $N_d$ dipoles:
\begin{equation}
    \lambda_{d}^2= -\frac{N_d}{4\pi^2 \rho^2_d}\log\vert z_{XY}\vert^2
    \label{eqn:dipole_corr_mb}
\end{equation}
whre $z_{XY}=\langle \Psi_{N_d}\vert U_{XY}\vert\Psi_{N_d}\rangle$ and $\rho_d=\frac{N_d}{L_xL_y}$ being the dipole density. Most of the models considered in the following section we will have exactly one dipole per unit cell meaning that $\rho_d=1/a^2$ with $a$ being the lattice constant in both directions. Also note that the equality (\ref{eqn:dipole_corr_mb}) is strict only in the thermodynamic limit: $N_d\to\infty$, $L_x,L_y\to\infty$ with $\rho_d=const$. One can also derive the localization area from the many-body localization tensor that can be related to the quantum metric over the space of rank-2 twists. We show this and provide an example context toward the end of Section \ref{sec:2Ddipoleinsulator}.

To complete the analogy to the charge case, let us now relate the dipole conductivity $\sigma_d$ to the magnitude of $U_{XY},$ and hence to the quadrupole fluctuations through a fluctuation dissipation theorem. For a finite system with open boundaries, the \textit{total} dipole current for a Hamiltonian $\mathcal{H}$ can be written as $J_{xy}^d = (i/\hbar)[\mathcal{H}, \widehat{XY} ]$ \cite{wheeler2018many} (we note that we set electric charge to unity throughout). %As we have shown we have shown in Appendix @@@@ we can define a dipole  $J_{xy}^d = \sigma_d E_{xy}$, where $E_{xy} = i\omega A_{xy}(\omega)$.  
As shown in App. \ref{app:linearresponse}, the real part of the dipole conductivity can be expressed via standard linear response calculations as
\begin{eqnarray}
\text{Re}\, \sigma_d(\omega)  =&&\frac{\pi}{V \hbar \omega}\sum_{n\neq 0}\bra{0}J_{xy}^d\ket{n}\bra{n}J_{xy}^d\ket{0} \nonumber\\ && \times[\delta(\omega_n - \omega) - \delta(\omega_n + \omega)],
\end{eqnarray}
where $\vert n\rangle$ is an eigenstate of H, and $\hbar \omega_n = E_n - E_0$. Using the commutation relation above we have $\bra{0} J_{xy}^d \ket{n} = -\omega_n \bra{0} \widehat{XY} \ket{n}$, and we can rewrite the integral of the real part of the conductivity in terms of $\widehat{XY}$ as 
\begin{equation}
\begin{split} \frac{V\hbar}{\pi} \int^\infty_0  \frac{d\omega \text{Re}\, \sigma_d(\omega)}{\omega} =  \sum_{n\neq 0} \bra{0}\widehat{XY}\ket{n}\bra{n}\widehat{XY}\ket{0}.
\end{split}
\end{equation}
Finally, we will make use of the fact that the energy eigenstates $\ket{n}$ form a complete set, and that we can replace $\sum_{n\neq 0} \ket{n}\bra{n}$ with $1 - \ket{0}\bra{0}$. Using this, we can express the real part of the conductivity as
\beq
\frac{V\hbar}{\pi} \int^\infty_0 \frac{d \omega \text{Re}\, \sigma_d(\omega)}{\omega} = \langle \widehat{XY}\widehat{XY}\rangle_0 -\langle \widehat{XY} \rangle_{0}^2, 
\eeq
which is the fluctuation dissipation theorem for the dipole conductivity. This result can be applied to periodic systems by making use of Eq. \ref{eq:quadfluct}. 

Now, after the development of this group of analogies between charge localization and dipole localization, we can provide a criterion to distinguish dipole insulators and dipole metals, given that the system is already a charge insulator (and has a conserved dipole moment). The criterion is just whether or not the ground state has delocalized dipoles. For dipole insulators $\lambda_d$ is finite, or equivalently, $|z_{XY}| \rightarrow 1$ as the system size becomes infinite. In dipole metals we have $\lambda_d\to\infty$, $\vert z_{XY}\vert\to 0,$ and the quadrupole fluctuations diverge as we approach the thermodynamic limit. Interestingly it seems there are two ways for a dipole insulator to delocalize: (i) the quadrupole fluctuations can diverge and the system will become a dipole metal, while remaining a charge insulator; (ii) if we lose exact dipole conservation the dipole fluctuations could become finite and eventually diverge if the system becomes an ordinary charge metal.

Let us now provide intuition for our criterion using two essentially classical examples. First we will present the results for a localized dipole probability density, and then we will show results for an extended dipole wave configuration.

\subsection{Localized Dipole}
 Consider the localized dipole  probability density
\beq
|\Psi(\bm{x}_1,\bm{x}_2)|^2 = \frac{1}{\pi\sigma^2}\delta(\bm{x}_1 - \bm{x}_2 - \bm{d}) e^{-|\bm{x}_1 + \bm{x}_2-2\bm{R}|^2 / 4 \sigma^2}.
\eeq
This probability distribution describes, e.g.,  an electron at $\bm{x}_1$ and a hole at $\bm{x}_2$ separated by a fixed (dipole) vector $\bm{d},$ and where the center of mass of the dipole is Gaussian localized near $\bm{R}$  with a variance of $\sigma^2$.  This wavefunction is an eigenstate of of $U_X$ and $U_Y$ with eigenvalues, $e^{2\pi i d_x/L_x}$ and $e^{2\pi i d_y/L_y}$ respectively, and we can identify $\bm{d}$ as the total dipole of the system. For $\sigma \ll L_x, L_y$, we find to leading order
\begin{eqnarray}
 z_{XY}  &=& \exp\left[\frac{-\pi^2|\bm{d}|^2 \sigma^2}{ L_x^2 L_y^2}\right]\exp\left[2 \pi i q_{xy}\right],\nonumber\\
\lambda_d &=& |\bm{d}| \sigma/\sqrt{2},\\ q_{xy}&=&(d_yR_x + d_xR_y)/L_x L_y.\nonumber
\end{eqnarray} In the thermodynamic limit $L_x L_y\gg \sigma^2, |\bm{d}|^2$ we find that the magnitude $|z_{XY}|\to 1,$ and thus, this configuration would be classified as a localized dipole state. Interestingly, we find that $\lambda_d$ can diverge if either $\sigma$ or $|\bm{d}|$ becomes large, i.e., if the dipole can move around freely, or if the dipole itself becomes unbound. We note in passing that the $q_{xy}$ calculated here is not independent of the choice of origin since we only consider a single dipole and the total dipole moment of the system is non-vanishing. 

\subsection{Dipole wave}
To provide an example of a system hosting  dipole currents, let us use a simple version of the rank-2 Lagrangian density from Ref. \onlinecite{pretko2018gauge} that contains only $xy$ kinetic terms for dipoles, similar to what  we have been considering above:
\begin{equation}
    \mathcal{L}=|\partial_t \Phi|^2-m^2|\Phi|^2-\lambda|\Phi\partial_x\partial_y\Phi-\partial_x\Phi\partial_y\Phi|^2.
\end{equation} Performing a Legendre transform, we arrive at the Hamiltonian density:
\begin{equation}
    \mathcal{H}=|\partial_t \Phi|^2+m^2|\Phi|^2+\lambda|\Phi\partial_x\partial_y\Phi-\partial_x\Phi\partial_y\Phi|^2
\end{equation}
This Hamiltonian density itself is sensitive to the action of the $U_{XY}(\mathfrak{q})$ twist:
\begin{equation}
    \begin{split}
        \mathcal{H}&(\mathfrak{q})=U_{XY}^{-1}(\mathfrak{q})\mathcal{H} U_{XY}(\mathfrak{q})=\\
        &=|\partial_t \Phi|^2+m^2|\Phi|^2+\lambda|\Phi\partial_x\partial_y\Phi-\partial_x\Phi\partial_y\Phi-i\mathfrak{q}\Phi^2|^2.
    \end{split}
\end{equation}
Using a class of classical `dipole wave' solutions (see App. \ref{app:dipolewave} for more detail) \begin{equation}
    \Phi_{\mathfrak{p}}(t,x,y)=\varepsilon\text{e}^{i\left(\mathfrak{p}xy-\omega t\right)}
\end{equation}
we can calculate both the dipole current (\ref{eq:dipolecurrent}) and the dipole stiffness (\ref{eq:dipolestiffness}) at the twist value $\mathfrak{q}$: 
\begin{equation}
    \begin{split}
        j_d(\mathfrak{q})=&\frac{\lambda}{V}\int dxdy \Big[iD_{xy}[\Phi_{\mathfrak{p}}]^\dagger \Phi_{\mathfrak{p}}^2-i(\Phi_{\mathfrak{p}}^\dagger)^2D_{xy}[\Phi_{\mathfrak{p}}]\\
        &+2\mathfrak{q}|\Phi_{\mathfrak{p}}|^4\Big]
        =2\lambda\varepsilon^4(\mathfrak{p}+\mathfrak{q})
    \end{split}
    \label{eqn:dipole_current}
\end{equation}
\begin{equation}
    D_d=2\frac{\pi\lambda}{V}\int dxdy|\Phi_{\mathfrak{p}}|^4=2\pi\varepsilon^4\lambda.
    \label{eqn:dipole_stiff}
\end{equation}
Both of these quantities are non-zero. On the other hand, using the twist operator method to calculate regular charge current and Drude weight for this Hamiltonian trivially gives zero meaning that the system at hand is a conventional insulator. So $\mathcal{H}$ along with the elementary field configuration $\Phi_\mathfrak{p}$ does indeed describe a well-defined dipole-conducting system.

\section{Lattice models for 1D dipole metals and insulators}
As was mentioned in the previous section, dipole conservation manifests in an invariance under symmetry transformations of the form $\text{e}^{i\boldsymbol{\alpha}\cdot \textbf{x}}.$ Let us consider how that restricts the form of possible lattice Hamiltonians. A generic term in a Hamiltonian in a second-quantized language is proportional to $c^\dagger_{i_1}...c^\dagger_{i_n}c_{j_1}...c_{j_m}$. As is well known, an invariance with respect to global phase rotations $c^\dagger_i\to \text{e}^{-i\theta}c^\dagger_i$, $c_i\to \text{e}^{i\theta}c_i$ allows only terms with the same number of creation and annihilation operators, e.g., a quadratic hopping term $c^\dagger_i c_j$. As a result, any Hamiltonian that respects a global $U(1)$ charge-conservation symmetry, conserves total charge in the system. 

Requiring  an additional invariance with respect to linearly varying phase rotations $\text{e}^{i\boldsymbol{\alpha}\cdot \textbf{x}}$, which we call a dipole $U(1)$ symmetry, places restrictions on the relative coordinates of the operators that can enter the Hamiltonian. For example for an operator 
\begin{equation}
c^\dagger(x^+_1,y^+_1)...c^\dagger(x^+_n,y^+_n)c(x^-_1,y^-_1)...c(x^-_n,y^-_n)
\end{equation}
to be invariant under both $\text{e}^{i\alpha x}$ and $\text{e}^{i\beta y}$ phase rotations we need to satisfy the following two constraints on the coordinates of the creation and annihilation operators: $\alpha\left(\sum_{i=1}^n x^+_i-\sum_{i=1}^n x^-_i\right)=2\pi k$ and $\beta\left(\sum_{i=1}^n y^+_i-\sum_{i=1}^n y^-_i\right)=2\pi l.$ Since this must be true for any real values of $\alpha$ and $\beta$ and some integers $k$ and $l$ this requires the sum of $x$ and $y$ coordinates of all annihilated electrons to equal to the sum of $x$ or $y$ coordinates of all created electrons respectively. Hence,  any term in the Hamiltonian allowed by both charge and dipole $U(1)$ symmetries, must conserve both total charge and dipole moment of the system. An elementary example of the type of term that satisfies both of these constraints is a dipole hopping term: 
\begin{equation}
\begin{split}
d_y^\dagger&(x,y)d_y(x+1,y)=\\
&c^\dagger(x,y+1)c(x,y)c^\dagger(x+1,y)c(x+1,y+1).
\end{split}
\end{equation}
We will use quartic terms like these as basic building blocks to construct our lattice models. 

\subsection{1D dipole metal, dipole insulator, and dipole superconductor}
\label{sec:dipole_metal}

Let us consider a two-leg ladder of fermion orbitals with its length along the $x$-direction, and a Hamiltonian with nearest-neighbor \emph{dipole} hopping interactions and onsite Hubbard repulsion:
\begin{equation}
    H=\frac{J}{2}\sum_{i=1}^N(d^{\dagger}_i d^{\phantom{\dagger}}_{i+1}+h.c.)+U\sum_{i=1}^N n_{i\uparrow}n_{i\downarrow},
    \label{eqn:dipole_metal}
\end{equation}
where $\uparrow/\downarrow$ label the two legs of the ladder, $d_i\equiv d^y_i\equiv c^\dagger_{i\downarrow}c_{i\uparrow}$ is a dipole annihilation operator for a dipole oriented perpendicular to $x$, i.e., pointing along the rungs of the ladder, and $c^{\dagger}_{i\downarrow/\uparrow}$ creates a fermion on the lower/upper row respectively. Thus, the $d_i$ operator annihilates a particle at site $i$ in the upper row and creates a particle on the lower row.
The dipole operators commute on different sites:
\begin{equation}
    [d^\dagger_i,d^\dagger_j]=[d_i,d_j]=[d^\dagger_i,d_j]=0,\ i\neq j,
    \label{eqn:dipole_commute}
\end{equation}
while operators belonging to the same site obey
\begin{equation}
    \{d^\dagger_i,d_i\}=n_{i\uparrow}+n_{i\downarrow}-2n_{i\uparrow}n_{i\downarrow}.
    \label{eqn:dipole_anticommute}
\end{equation}

\begin{figure}
	\includegraphics[width=0.45\textwidth]{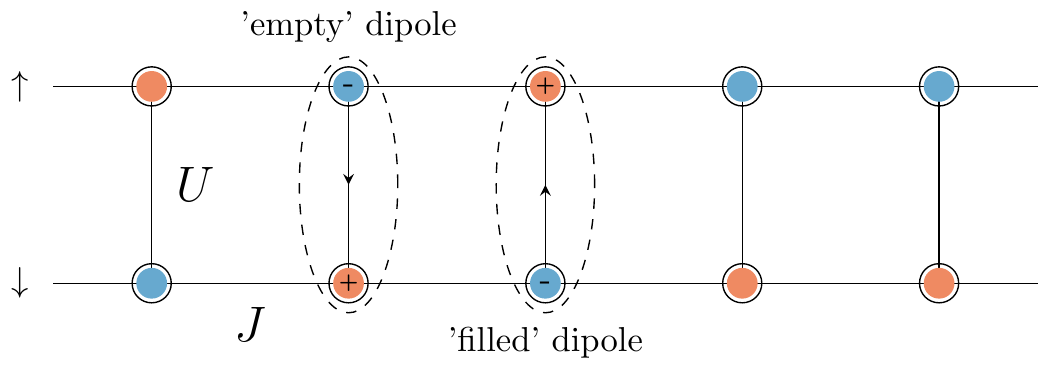}
    \caption{A configuration of a 1D dipole model state. Electrons and holes are depicted by blue and red circles respectively. Strong value of the interaction $U$ ensures that there is exactly one positive and one negative charge per rung of the ladder.}
\label{fig:dipole_ladder}
\end{figure}

Now, let us restrict our model to be at a half-filling; since there is one fermion orbital per row, per site,  half-filling means $N_x$ fermions. Additionally we are going to take the potential $U\gg J$ to guarantee exactly one fermion per rung of the ladder, meaning that on every rung $i$  we have a well-defined dipole state - either empty or occupied as determined by $n^d_i\equiv d^{\dagger}_i d_i=n_{i\uparrow}(1-n_{i\downarrow})$. 
%\textbf{OSD: I think that the sentence above correctly captures the reason for the constraint (to have a well-defined dipole state). Below I added a couple of sentences clarifying it.}
Without this constraint the dipole occupation number would be ill-defined, as the state $c^\dagger_{i\uparrow}c^\dagger_{i\downarrow}\ket{0}$ is annihilated by both $d_i$ and $d^\dagger_i$. \footnote{As a consequence, if such states were allowed, we could no longer relate a dipole particle number $N^d=\sum_i d^\dagger_i d_i$ to a physical polarization: a pair of states $c^\dagger_{i\uparrow}c^\dagger_{i\downarrow}\ket{0}$ and $c^\dagger_{i\uparrow}c^\dagger_{i+1\downarrow}\ket{0}$ has the same polarization in the direction transverse to the ladder but different dipole numbers: for the first one $N^d=0$, for the second one $N^d=1$.}
This constraint further simplifies the on-site anti-commutation relation (\ref{eqn:dipole_anticommute}) to be:
\begin{equation}
    \{d^\dagger_i,d_i\}=1,\ \{d^\dagger_i,d^\dagger_i\}=\{d_i,d_i\}=0,
    \label{eqn:dipole_anticommute2}
\end{equation}
which means that we can now interpret the dipoles $d_i$ as conventional hard-core bosons. 

If the dipoles are effectively hard-core bosons, our model can be equivalently rewritten as a spin-1/2 XY model via:
\begin{equation}
    S^\alpha_i=\frac12 \vec{c}_{i}^{\ \dagger} \sigma^\alpha \vec{c}_{i},\ \text{where}\  \vec{c}_{i}=(c_{i,\uparrow},c_{i\downarrow})^T,
    \label{eqn:fermions_to_spin}
\end{equation}
so that
\begin{equation}
    d^\dagger_i=2S^+_i,\ d_i=2S^-_i.
    \label{eqn:dipoles_to_spin}
\end{equation}
The resulting spin Hamiltonian is
\begin{equation}
    H=2J\sum_{i=1}^N \left(S^+_iS^-_{i+1}+S^-_{i}S^+_{i+1}\right).
    \label{eqn:heisenberg_model}
\end{equation}
This model is exactly solvable in 1D as we can map it to a free-fermion model via a Jordan-Wigner transformation:
\begin{equation}
    S^+_i=\text{e}^{i\pi\sum_{j=1}^{i-1}c^\dagger_j c_j}c^\dagger_i,\ S^-_i=\text{e}^{-i\pi\sum_{j=1}^{i-1}c^\dagger_j c_j}c_i,
    \label{eqn:jw-map}
\end{equation}
and the resulting transformed Hamiltonian is:
\begin{equation}
    H=2J\left(\sum_{i=1}^{N-1}c^\dagger_i c_{i+1}+\text{e}^{i\pi\sum_{j=1}^{N}c^\dagger_j c_j}c^\dagger_N c_1\right)+h.c.
    \label{eqn:fermion_ham}
\end{equation}

Now we would like to understand the properties of the ground state of this Hamiltonian. To show that the ground state of this model is a \emph{dipole} metal we first need to prove that the model  (\ref{eqn:dipole_metal}) describes a conventional \emph{charge} insulator. Following Kohn's work \cite{kohn1964} we demonstrate that this system is insensitive to twisted boundary conditions. By placing our dipole hopping Hamiltonian in a constant external (charge) gauge field $\textbf{A}(x)=(k_x,k_y)$, and coupling to the lattice hopping terms with Peierls factors, we find every single term in the Hamiltonian is left completely unchanged. Notably, the dipole hopping terms would only couple only to a gradient of the charge gauge field. Therefore, the Hamiltonian does not acquire any dependence on $k_x$ or $k_y,$ and hence the corresponding values of the charge stiffnesses trivially vanish:
\begin{equation}
    D_{c}^{\alpha}\equiv-\frac{\pi e^2}{V}\frac{\partial^2 E}{\partial k_\alpha^2}=0.
    \label{eqn:vanishing_drude}
\end{equation}

Hence, introducing constant charge gauge field in the dipole case doesn't have any effect.
On the other hand, by introducing a constant gauge field gradient $\textbf{A}(x)=(\mathfrak{q}x,\mathfrak{q}y)$ we find that every term picks up a phase factor
\begin{equation}
    d^{\dagger}_i d_{i+1}\to \text{e}^{i\mathfrak{q}}d^{\dagger}_i d_{i+1}.
    \label{eqn:dipole_hopping_phase_factor}
\end{equation}
By taking a derivative of $H$ with respect to $\mathfrak{q}$ we obtain a dipole current operator:
\begin{equation}
    J_d(\mathfrak{q})=i\frac{J}{2}\sum_i^N\left(\text{e}^{i\mathfrak{q}}d^{\dagger}_i d^{\phantom{\dagger}}_{i+1}-\text{e}^{-i\mathfrak{q}}d^{\dagger}_{i+1} d^{\phantom{\dagger}}_{i}\right).
\end{equation}
As we can see, the combination $\frac{1}{2}(\partial_x A_y+\partial_y A_x)=\mathfrak{q}$ plays the role of the gauge field for our dipole model and so it is natural to treat it as a rank-2 gauge field (see Appendix \ref{app:rank1rank2})\footnote{While this replacement seems innocuous the behaviour at the boundary is the crucial difference between the rank-1 and rank-2 descriptions. A rank-2 field can be constant everywhere on a periodic manifold while derivatives of a rank-1 field cannot. This is similar to the fact that  on a 1D ring we can have a constant rank-1 field, but there's no such smooth configuration of potential field $\phi(x)$ such that $A_x=\partial_x\phi(x)$ is constant everywhere. This suggests that in these arguments it may be important to consider coupling to rank-2 field instead of a derivative of a rank-1 field, though it may be possible to fix the issue in the rank-1 case.} 
\begin{equation}
    A_{xy}\equiv\frac{1}{2}(\partial_x A_y+\partial_y A_x).
\end{equation}
%or, equivalently, a constant rank-2 gauge field {\bf{TLH: Need to discuss how this model can couple to a rank-2 gauge field; this should be connected to the dipole conservation we list earlier.}} $A_{xy}=\frac{1}{2}(\partial_x A_y+\partial_y A_x)=\mathfrak{q}$ across the chain, we attach the same phase factor to every dipole hopping term in the Hamiltonian (\ref{eqn:dipole_metal}):

Now we can take this modification to the dipole Hamiltonian and push it through to the fermionic model after the spin-mapping and subsequent Jordan-Wigner transform. By doing so we arrive at a fermionic model of the form Eq. (\ref{eqn:fermion_ham}), but where every term multiplied by the same phase factor $\text{e}^{i\mathfrak{q}}$. Thus, a rank-2 twist of the dipole Hamiltonian results in a conventional rank-1 twist of the fermion model. The twisted fermion model can be solved at any given filling, and the ground state energy is always dependent on $\mathfrak{q}$, (as long as the states are not completely filled or empty) so the dipole stiffness (dipole Drude weight) (\ref{eqn:ddw1_expression}) is \emph{non-vanishing}:
\begin{equation}
    D_d=-\frac{\pi}{V}\left.\frac{\partial^2 E}{\partial \mathfrak{q}^2}\right|_{\mathfrak{q}=0}>0.
    \label{eqn:non-vanish_ddw}
\end{equation} Hence this system represents a dipole metal.

Remarkably we have found that placing a 1D dipole chain into a constant \emph{gradient} of the charge gauge field is mapped to an ordinary fermionic chain in a constant gauge field. The gapless fermionic theory that carries a charge current is interpreted in the boson language as a theory having charge-neutral quasiparticles that carry dipole moment and respond to a rank-2 gauge field (or a essentially equivalently, responds to gradients of the rank-1 electric field). 

\emph{Dipole insulator.} Having a direct map of the dipole metal Hamiltonian (\ref{eqn:dipole_metal}) to the fermionic chain (\ref{eqn:fermion_ham}), we can immediately provide an example of a 1D dipole insulator. If we dimerize the dipole hopping term strengths then our dipole model maps to an ordinary fermionic Su-Schrieffer-Heeger (SSH) chain\cite{SSH1979} after the Jordan-Wigner transformation. As we established, there is a direct equivalence between the dipole stiffness $D_d$ of the 1D dipole model and the charge stiffness of the fermionic chain to which the dipole model is mapped. Therefore, the dipole stiffness of our 1D dipole chain with alternating couplings is vanishing in the thermodynamic limit as the charge stiffness in the SSH chain vanishes in that limit.

\textit{Twist Operators.} Now let us confirm these results using the twist operator approach to localization. First, we note that every term in the dipole Hamiltonian (\ref{eqn:dipole_metal}) commutes with both the $U_X$ and $U_Y$ operators, so the ground state of the dipole model can be written as an exact eigenstate of these operators. Thus we have $|z_X|=|z_Y|=1.$ To analyze the expectation value of the $U_{XY}$ operator, let us first assign the $y$-coordinate to be $+1/2$ and $-1/2$  for the top and bottom chains respectively in our two leg ladder (\ref{eqn:dipole_metal}) (changing these values just changes $U_{XY}$ below by an unimportant constant phase factor). This allows us to rewrite $U_{XY}$ as:
\begin{equation}
    \begin{split}
        U_{XY}&=\exp\left(i\frac{2\pi}{L_xL_y}\sum_{x,y=1}^{L_x,L_y} xy\hat{n}_{x,y}\right)\\
        &=\exp\left(i\frac{2\pi}{N}\sum_{x=1}^{N} x\frac12(\hat{n}_{x\uparrow}-\hat{n}_{x\downarrow})\right).
    \end{split}
\end{equation}
Note that we effectively have doubled the unit cell in the $\hat{y}$ direction which allowed us to put $L_y=1$. We can rewrite this operator in a spin basis via (\ref{eqn:fermions_to_spin}) to find: 
\begin{equation}
    U_{XY}=\exp\left(i\frac{2\pi}{N}\sum_{j=1}^NjS^z_j\right),
\end{equation}
which is just a conventional $U_X$ twist operator for a 1D spin chain. Under the Jordan-Wigner map (\ref{eqn:jw-map}) this operator simply transforms into the $U_X$ operator for the resulting fermionic chain (\ref{eqn:fermion_ham}). Thus, the expectation value $z_{XY}$ in the $N_x\to\infty$ limit can be evaluated in the fermionic language to be $z_{XY}\approx 0$ in the dipole metal phase, and $z_{XY}=\exp(i P)\left[\cos(\pi/N_x)\right]^{N_x}\approx\pm 1$ in the insulating case where the couplings are dimerized. The absolute value of $z_{XY}$ allows us to calculate dipole correlation area as a function of the length of the ladder: 
\begin{equation}
    \lambda^2_d=-\frac{N_d}{4\pi^2\rho_d^2}\log \left[\cos^{2N_x}\left(\frac{\pi}{N_x}\right)\right]\approx \frac{a^4}{4}+O\left(\frac{1}{N_x}\right),
    \label{eqn:2leg_corr}
\end{equation}
which gives $\lambda_d=a^2/2$ in the zero-correlation length limit.
The phase $P$ takes values such that there is a relative phase of $\pi$ between the dipole insulator models having the intra- or inter-cell dipole hopping terms dominate. $P$ is quantized because of the inversion symmetry of the model, and would be interpreted as a quantized charge polarization in the fermionic model. Let us see how we can interpret this phase in the dipole language.

\textit{Quadrupole from Berry phase} 
It is straightforward now to introduce the notion of Berry phase with respect to the parameter $\mathfrak{q}$. As was already discussed, the dipole hopping Hamiltonian translates to a regular fermionic chain with the parameter $\mathfrak{q}$ entering as a parameter for the Berry phase that is used to calculate the polarization of the 1D fermionic chain. Therefore, when the  dipole hoppings are dimerized the Berry phase computed for the ground state $|\Psi(\mathfrak{q})\rangle$ of the dipole Hamiltonian:
\begin{equation}
   \text{Im}\int_{0}^{\frac{2\pi}{N}} d\mathfrak{q}\langle\Psi(\mathfrak{q})|\partial_{\mathfrak{q}}|\Psi(\mathfrak{q})\rangle=P
\end{equation}
where $P$, once again, takes such values that there is a relative $\pi$ phase between the Berry phases computed for the ground state of two Hamiltonians with opposite dimerization patterns. We have already related this Berry phase to the quadrupole moment $Q_{xy}$, hence this system has a quantized quadrupole moment.

\textit{Dipole Superconductor.}
Finally, we can push this model further to consider a dipole superconductor by adding an on-site chemical potential for dipoles $\mu_d,$ and a p-wave \emph{dipole pairing} term with a strength $\Delta$:
\begin{equation}
    H=\sum_{i=1}^N \left(\frac{J}{2}d^{\dagger}_i d_{i+1}+\Delta d^\dagger_id^\dagger_{i+1}+h.c.\right)-\mu_d \sum_id^\dagger_i d_i.\label{eq:dipsc}
\end{equation}
The pairing terms break the $U(1)$ global dipole conservation down to $\mathbb{Z}_2$. Using our mapping, this model yields a 1D Kitaev chain\cite{kitaev2001} The model exhibits a weak pairing phase, corresponding to a topological superconductor, occurring when $|\mu_d|<J$ and a a strong pairing - trivial superconducting phase - appearing when $|\mu_d|>J$. Hence, this model represents a gapped dipole superconductor when $\mu_d\neq 0.$ Since the fermion Kitaev chain has a non-vanishing charge stiffness $D_c,$ we find that the dipole superconductor (\ref{eq:dipsc}) has a non-vanishing dipole stiffness $D_d.$ If we recall the result of Ref. \onlinecite{scalapino1993}, we can conclude that, since the Kitaev chain is gapped, it also has a non-vanishing (charge) superfluid density $D_{c,s}.$ Thus we can claim that there should be an analogous dipole superfluid density $D_{d,s}.$ We leave a full discussion of the connection between the dipole superfluid density and the dipole stiffness to future work. We also note that a dipole superconductor/supefluid has been proposed in bilayer systems, and it would be interesting to compare with the discussion here in future work\cite{eisenstein2004,jiang2015}.

\subsection{Realization of a 1D dipole metal and dipole insulator in a cold atom system}
Let us now provide a method for constructing a  dipole metal and dipole insulator analogous to Eq. \ref{eqn:dipole_metal} in a 1D cold-atom context. To model a system of ultra cold bosonic atoms, we will consider an extended two-leg Bose-Hubbard ladder with two spin states, of the form:
\begin{equation}
\begin{split}
 H_{\text{BHM}} = &-\sum_{j,i,\sigma = \uparrow,\downarrow} t(a^\dagger_{j,i,\sigma}a_{j+1,i,\sigma} + h.c.)\\   &- \sum_{j,\sigma} t' (a^\dagger_{j,1,\sigma}a_{j,2,\sigma}  + h.c.)\\  &+ \sum_{j,i} U [n_{j,i}(n_{j,i}-1)] \\ &+ \sum_{j} V(n_{j,1}n_{j,2})
 \\ &+ \sum_{j}\frac{\Delta_{i}}{2} (n_{j,i,\uparrow}-n_{j,i,\downarrow}) + \mu\sum{j,i} n_{j,i}.
\label{eq:BHM}
\end{split}  
\end{equation}
 Here $a^{\dagger}_{j,i,\sigma}(a_{j,i,\sigma})$ is the creation (annihilation) operator for a boson of spin $\sigma$ on rung $j$ and leg $i = 1,2$ of the ladder, and $n_{j,i} = n_{j,i \uparrow}+n_{j,i \downarrow}$. $U$ is the on site interaction, $V$ is the nearest neighbor interaction between bosons on different legs of the ladder, $\Delta_{i}$ is a spatially varying magnetic field, and $\mu$ is the chemical potential. %We are assuming that the distance between rungs is great enough such that the interactions between bosons on different rungs can be ignored 
 
Here we consider the system at quarter filling, and where $U$ is significantly strong enough such that there is one boson per site, and the bosons do not condense. Additionally, we will set $\Delta_{i}$  such that $\Delta_{1} < 0$ and $\Delta_{2} > 0$, and $|\Delta_{1}|, |\Delta_{2}| \gg t$, i.e., there is an effective magnetic field gradient along the ladder rungs. This will suppress the $t'$ hopping along rungs, and at low energy will confine the spin up bosons to leg 1 and the spin down bosons to leg 2. Using this, we can reduce the Hamiltonian Eq. \ref{eq:BHM} to
\beq
\nonumber H'_{\text{BHM}} &=& -\sum_{j} t(a^\dagger_{j,1,\uparrow}a_{j+1,1,\uparrow} + h.c.)\\\nonumber  &\phantom{=}& -\sum_{j} t(a^\dagger_{j,2,\downarrow}a_{j+1,1,\downarrow} + h.c.)\\\nonumber  &\phantom{=}& + \sum_{j} Vn_{j,1,\uparrow}n_{j,2,\downarrow}\\  &\phantom{=}& + \sum_{j} [\Delta_{1} n_{j,i,\uparrow}-\Delta_{2}n_{j,i,\downarrow}].
\label{eq:BHM2}
\eeq
If we suppress the (now redundant) leg index, we find  exactly the Hamiltonian for a single Bose-Hubbard model with two spin states, at half filling. One important feature of this construction is that in Eq. \ref{eq:BHM2}, the bosons of different spins are also located at different locations in space, i.e., on the different legs of the ladder. In the limit of large $V$, Eq. \ref{eq:BHM2} becomes the $XXZ$ model \cite{duan2003,trotzky2008} 
\beq
\nonumber H_{\text{XXZ}} &=& J \sum_j (\frac{1}{2}[S^+_j S^-_{j+1} + S^-_j S^+_{j+1}] - S^z_j S^z_{j+1})\\ &\phantom{=}& + \sum_j \bar{\Delta}S^z_j
\label{eq:XXZ}
\eeq
where $\bar{\Delta} = \Delta_{1} + \Delta_{2}$ is the average magnetic field of the system. The spin creation operator $S^+_j = a^\dagger_{j,1,\uparrow}a_{j,2,\downarrow}$ can also be interpreted as a \emph{dipole} creation operator, because of the spatial separation between the up and down spin bosons on the ladder. The operator $S^z_j = (n_{j,1,\uparrow} - n_{j,1,\downarrow})/2$ can be interpreted as the `dipole occupancy' of site $j$. 

Eq. \ref{eq:XXZ} resembles the dipole ladder Eq. \ref{eqn:dipole_metal} with the addition of a field $\Delta$, and the additional nearest neighbor dipole interaction term $S^z_j S^z_{j+1}$. By use of an appropriate unitary transformation, the XXZ model in Eq. \ref{eq:XXZ} can be mapped onto a ferromagnetic Heisenberg spin-1/2 chain. The spectrum of the ferromagnetic Heisenberg spin-1/2 chain consists of gapless spin waves, which correspond to gapless dipolar excitations in Eq. \ref{eq:XXZ}. This indicates that Eq. \ref{eq:XXZ} describes a dipole metal. 

We can also consider a variation of this system where we arrange the ladder rungs such that the boson hoppings between $j$ and $j+1$ for odd values of  $j$ are suppressed. In this limit the dipoles dimerize, and the resulting dipole Hamiltonian becomes,
\beq
\nonumber H_{\text{XXZ}} &=& - J \sum'_j (S^+_j S^-_{j+1} + S^-_j S^+_{j+1} - S^z_j S^z_{j+1})\\ &\phantom{=}& + \sum_j \bar{\Delta}S^z_j ,
\label{eq:XXZ2}
\eeq
in the spin language, where the $'$ indicates a sum only over even $j$. Due to the dimerization of Eq. \ref{eq:XXZ2}, all excitations are localized. Each dimer will be in a triplet configuration of the two effective spin-1/2s and the spectrum will be gapped for $|\bar{\Delta}|   > 0$. Thus we expect this to represent a dipole insulator.

\section{2D dipole insulator and Dipole Metal}\label{sec:2Ddipoleinsulator}
\begin{figure}
	\includegraphics[width=0.45\textwidth]{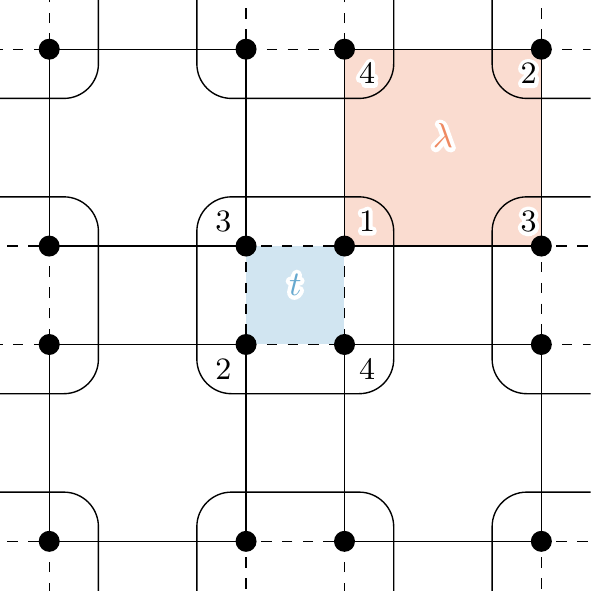}
    \caption{Model for a 2D dipole insulator having quartic ring-exchange couplings on alternating plaquettes (solid and dotted lines) with alternating couplings ($\lambda$ and $t$).}
\label{fig:quadrupole-tb}
\end{figure}

Now we will move on to a discussion of a model for a 2D dipole insulator. We start with a square lattice model with four degrees of freedom per unit cell. These degrees of freedom can be fermions or hardcore bosons, but to be explicit let us choose fermions. The model we consider is a model with plaquette ``ring-exchange" couplings inside a unit cell or between unit cells (see Fig. \ref{fig:quadrupole-tb}). A bosonic version of this model was recently considered in Ref. \onlinecite{ybh2019} where it was argued that it could represent a topological dipole insulator having a quantized quadrupole moment. Explicitly the model Hamiltonian consists of quartic interacting terms of the following form:
\begin{equation}
    \begin{split}
    H=\lambda\sum_{\textbf{p}}\left|\swne\right\rangle\left\langle\nwse\right|+t\sum_{\textbf{s}} \left|\swne\right\rangle\left\langle\nwse\right|+h.c.
    \end{split}
    \label{eqn:Ham4pt}
\end{equation}
where $\tikz[baseline=-0.5ex]{\fill (0,0) circle (2pt);}$ depicts a site in an occupied state and $\tikz[baseline=-0.5ex]{\draw (0,0) circle (2pt);}$ depicts an empty site.
The first sum here runs over the interstitial plaquettes $\textbf{p}$ marked with the solid lines in Fig. \ref{fig:quadrupole-tb}, and the second sum runs over the on-site plaquettes $\textbf{s}$ marked with the dashed lines.
%\begin{equation}
%    \begin{split}
%    H=\lambda\sum_{\textbf{r}} c^\dagger_{\textbf{r},1}&c_{\textbf{r}+\hat{x},3}c^\dagger_{\textbf{r}+\hat{x}+\hat{y},2}c_{\textbf{r}+\hat{y},4}\\
%    &+t\sum_{\textbf{r}}c^\dagger_{\textbf{r},2} c_{\textbf{r},4} c^\dagger_{\textbf{r},1} c_{\textbf{r},3} +h.c.
%    \end{split}
%    \label{eqn:Ham4pt}
%\end{equation} 
%{\bf{TLH: Should we also write this in terms of some dipole hopping terms where we pick the dipoles to be oriented along x or y? OSD: I'm not sure if we should. We don't make use of the description in terms of dipoles later. I changed the description of the Hamiltonian to try to make it look more intuitive. Should we keep it?}}
%where the index $\textbf{r}$ runs over all unit cells. 
%The first collection of terms represent inter-cell couplings associated with plaquettes and are depicted as solid squares in Fig. \ref{fig:quadrupole-tb}, and the second collection of terms represents on-site couplings depicted by dashed squares. 
We will consider this model at half-filling. In the limit where $t=0,$ (or $\lambda=0$) we can actually solve this model exactly --all we need to do is to find the spectrum of the Hamiltonian at every plaquette. The ground state for any given plaquette is:
\begin{equation}
    \ket{\psi}_{\textbf{p}}
    %=\frac{1}{\sqrt{2}}\left(\ket{1100}_p-\ket{0011}_p\right)
    =\frac{1}{\sqrt{2}}\left(\left|\swne\right\rangle_{\textbf{p}}-\left|\nwse\right\rangle_{\textbf{p}}\right),
    \label{eqn:dipole_p_gs}
\end{equation}
%where the state $\ket{1100}_p,$ for example, has the sites labeled 1 and 2 in plaquette $p$ occupied. 
%Thus, the ground state for each plaquette is a linear combination of two states in which a pair of fermions is diagonally separated on the plaquette.
and the overall ground state for the lattice with periodic boundary conditions in both directions is then simply a tensor product:
\begin{equation}
    \ket{GS}=\bigotimes_{\textbf{p}}\ket{\psi}_{\textbf{p}}.
    \label{eqn:overall_gs}
\end{equation} We now want to show that this ground state is a dipole insulator.
%\noindent where $p$ runs over the interstitial plaquettes marked with the solid lines in Fig. \ref{fig:quadrupole-tb}.

\textit{Dipole stiffness.} First let us calculate the dipole stiffness by probing the energy of the ground state during the insertion of a constant rank-2 gauge field. For our model we can introduce minimal coupling to the rank-2 gauge field $A_{xy}$ through Peierls-like phase factors for the ring-exchange couplings.  Inserting a constant rank-2 gauge field $A_{xy}=\mathfrak{q}$ simply multiplies every term in the Hamiltonian by a phase factor $\text{e}^{i\mathfrak{q}}$. 

%\begin{figure}
%	\includegraphics[width=0.45\textwidth]{pics/perturbation.pdf}
 %   \caption{A single perturbation term $\hat{v}(\textbf{r})$ changes filling of the plaquettes adjacent to a cell at $\textbf{r}$.}
%\label{fig:perturbation}
%\end{figure}

Now we can compute the dipole stiffness for our model in the $t=0$ limit, and we immediately find a vanishing current  $J_d=0$ and dipole stiffness $D_d=0$ as the energy does not depend on the value of $\mathfrak{q}$ at all. Indeed, the situation is not changed after we turn on $t\neq 0$ as the on-site potential terms do not couple to the rank-2 gauge field, and so the perturbative corrections to the energy do not depend on $\mathfrak{q}$.\footnote{Even if one treats the degrees of freedom in the unit cell as being spatially separated within the cell, one can show that the first correction to the energy due to a non-vanishing $\mathfrak{q}$ is suppressed by a factor of $1/(\Delta E)^{L_x L_y}$ where $\Delta E$ is the bulk energy gap.}

\textit{Twist operators and Localization.} Our 2D Hamiltonian commutes with both $U_X$ and $U_Y$ operators and therefore, its eigenstates are also exact eigenstates of those twist operators. Thus we find $|z_X|=|z_Y|=1$, for the 2D dipole insulator model, and hence, from the localization criterion, we confirm that the system is a \emph{charge} insulator. On the other hand, $U_{XY}$ acts non-trivially on the terms in the Hamiltonian (\ref{eqn:Ham4pt}), and in case when $t=0$, $\lambda=1$ we can compute the expectation value of the dipole twist operator to be:
\begin{equation}
    \begin{split}
    z_{XY}&=\langle U_{XY}\rangle\\
    &=|z_{XY}|\exp\left(\frac{i\pi}{N_x N_y}(N_xN_y+1)(N_x+1)(N_y+1)\right) 
    %&\times \cos^{N_xN_y}\left(\frac{\pi}{N_xN_y}\right) \cos^{N_y}\left(\frac{\pi(N_x-1)}{N_xN_y}\right) \\
    %&\times \cos^{N_x}\left(\frac{\pi(N_y-1)}{N_xN_y}\right) \cos\left(\frac{\pi(N_x-1)(N_y-1)}{N_x N_y}\right)
    \end{split}
    \label{eqn:expectation_zxy}
\end{equation}
where the absolute value $|z_{XY}|\to 1$ in the thermodynamic limit. 
In the opposite case, when $t=1$, $\lambda=0$ we simply have:
\begin{equation}
    z_{XY}=\exp\left(i\pi(N_x+1)(N_y+1)\right).
\end{equation}
The relative phase between the values of $z_{XY}$ computed for these two cases in the thermodynamic limit, is exactly $\pi$. Furthermore, after we take into account the background ionic charge, we always find: $z_{XY}=-1$ when $t=0$, $\lambda=1$ and $z_{XY}=+1$ when $t=1$, $\lambda=0$. 
%quantities Phases of both of these quantities in a thermodynamic limit $N_x,N_y\to\infty$ take $\pm 1$ values depending on whether the product $N_x N_y$ is even or odd, however, the relative phase factor between the $z_{XY}$ computed for these two cases is always $-1$. 
As for the dipole correlation areas, in the limit with $t=1$, $\lambda=0$ we trivially obtain $\lambda_d=0$ as $|z_{XY}|=1$ exactly. In the opposite case we find (see Appendix \ref{app:zmagnitude}) the expectation value (\ref{eqn:expectation_zxy}) simplifies to: $|z_{XY}|^2=\left[\cos\left(\frac{\pi}{N_xN_y}\right)\right]^{2N_xN_y}$ and we have:
\begin{equation}
    \begin{split}
        \lambda^2_d=-\frac{N_d}{4\pi^2\rho_d^2}2N_x N_y & \log\cos\left(\frac{\pi}{N_xN_y}\right)\\
        &\approx \frac{a^4}{4} +O\left(\frac{1}{N_x N_y}\right),
    \end{split}
\end{equation}
from which we find: $\lambda_d=a^2/2$. This value of $\lambda_d$ coincides with the value we already computed for the two-leg ladder (\ref{eqn:2leg_corr}). This could have been anticipated as in both cases the system is built of a collection of local plaquette ring-exchange terms, so we would naturally expect to find the same answer for the dipole correlation areas in these two models.

Another way to reach the same result for the dipole correlation area is by introducing the quantum metric with respect to the rank-2 phase twist. In Refs. \onlinecite{marzari1997},\onlinecite{resta1999},  \onlinecite{souza2000}  it was shown that the quantum metric defined over the parameter space controlling the twisted boundary conditions can be related to the many-body localization tensor $\langle r_\alpha r_\beta \rangle_c$. Explicitly, boundary conditions along $x_\alpha$ can be twisted by threading a magnetic flux around the system which, in turn, can manifest  as a uniform gauge field $A_\alpha=\Phi_{\alpha}$ on the $\alpha$-links of the lattice. For any particular twist of the boundary conditions we can find the ground state of the corresponding Hamiltonian and denote it as $\ket{\Psi(\Phi_1,...\Phi_d)}=\ket{\Psi(\boldsymbol{\Phi})}$ The quantum metric on this parameter space is then defined as:
\begin{equation}
    \begin{split}
        g_{\alpha\beta}(\boldsymbol{\Phi})=\langle\partial_\alpha& \Psi(\boldsymbol{\Phi})|\partial_\beta \Psi(\boldsymbol{\Phi})\rangle\\
        &-\langle\partial_\alpha \Psi(\boldsymbol{\Phi})|\Psi(\boldsymbol{\Phi})\rangle\langle\Psi(\boldsymbol{\Phi})|\partial_\beta \Psi(\boldsymbol{\Phi})\rangle.
        \label{eqn:quant_metric}
    \end{split}
\end{equation}
In \onlinecite{Resta2005} this metric tensor was related the localization tensor in the system of $N$ electrons:
\begin{equation}
    \langle r_\alpha r_\beta \rangle_c=\frac{g_{\alpha\beta}(0)}{N}.
\end{equation}

Similarly, when working with  systems of dipoles, the uniform rank-2 gauge field $A_{ij}$ can serve as a twisted parameter space for dipoles across the $i-j$ plane. Introducing a set of rank-2 twists along all of the planes of the lattice we define the quantum metric exactly as in (\ref{eqn:quant_metric}) where the vector $\boldsymbol{\Phi}$ now parametrizes the rank-2 twists along planes of the lattice instead of coordinate directions: $A_{ij}=\Phi_{ij}$. We propose the following formula for the dipole localization tensor computed in the system of $N_d$ dipoles:
\begin{equation}
    \langle\widehat{r_\alpha r_\beta}\widehat{r_\delta r_\gamma}\rangle_c=\frac{g_{(\alpha\beta)(\delta\gamma)}(0)}{N_d},
\end{equation}
where the $g_{(\alpha\beta)(\delta\gamma)}$ is still a rank-2 metric tensor with two indices $(\alpha\beta)$ and $(\delta\gamma)$ which denote the corresponding planes. To test this formula, let us apply it to the ground state of our 2D dipole insulator model (\ref{eqn:Ham4pt}) in the limit where $\lambda=1$, $t=0$. The rank-2 twist $A_{xy}=\Phi$ will modify the ground state of each plaquette in the following way:
\begin{equation}
    \ket{\psi(\Phi)}_\textbf{p}=\frac{1}{\sqrt{2}}\left(\text{e}^{-\frac{i}{2}a^2\Phi}\left|\swne\right\rangle_{\textbf{p}}-\text{e}^{\frac{i}{2}a^2\Phi}\left|\nwse\right\rangle_{\textbf{p}}\right).
\end{equation}
The full ground state $\ket{\Psi(\Phi)}$ in this simple limit is given by a tensor product of $\ket{\psi(\Phi)}_\textbf{p}$ over all $N_x\times N_y$ plaquettes in the system. We can check that for any single plaquette we have $\langle\partial \psi(\Phi)|\psi(\Phi)\rangle_{\textbf{p}}=0$ and calculating the only component of the quantum metric is then simply:
\begin{equation}
    g_{(XY)(XY)}= \sum_{\textbf{p}}\langle\partial \psi(\Phi)|\partial \psi(\Phi)\rangle_{\textbf{p}}=N_x N_y\frac{a^4}{4}.
\end{equation}
After taking into account that the number of dipoles is $N_d=N_x\times N_y$, we find the following result for the dipole localization tensor:
\begin{equation}
    \langle\widehat{XY}\widehat{XY}\rangle_c=\frac{a^4}{4}.
\end{equation}
From here we recover the already familiar value of the dipole localization area for this particular system: $\lambda_d=a^2/2$.

\textit{Berry phase.} As in the previous section, we can introduce the Berry connection with respect to the parameter $\mathfrak{q}$ and compute it in two limiting cases: (i) $t=0$, $\lambda=1$ and (ii) $t=1$, $\lambda=0$. In both limits, the ground state is simply the tensor product of two-particle states living on disjoint plaquettes, and the Berry phase for the overall system is hence equal to the Berry phase computed for a single cluster. For the case where the on-site couplings $t$ dominate, the Hamiltonian does not couple to the rank-2 gauge field at all and the corresponding Berry connection and Berry phase trivially vanish. In the opposite limit we find that the ground state of a single plaquette $\textbf{p}$ depends on $\mathfrak{q}$:
\begin{equation}
    \ket{\psi(\mathfrak{q})}_{\textbf{p}}=\frac{1}{\sqrt{2}}\left(\text{e}^{-i\frac{\mathfrak{q}}{2}}\left|\swne\right\rangle_{\textbf{p}}-\text{e}^{i\frac{\mathfrak{q}}{2}}\left|\nwse\right\rangle_{\textbf{p}}\right),
    \label{eqn:dipole_p_gs_berry}
\end{equation}
and the Berry phase computed for a single plaquette is:
\begin{equation}
    \text{Im}\int_0^{2\pi}d\mathfrak{q}\langle\psi(\mathfrak{q})|\partial_{\mathfrak{q}}|\psi(\mathfrak{q})\rangle_\textbf{p}=\pi.
\end{equation}
When we consider the Berry phase associated with a periodic $L_x\times L_y$ lattice with the ground state $\ket{\Psi(\mathfrak{q})}=\bigotimes_{\textbf{p}}\ket{\psi(\mathfrak{q})}_{\textbf{p}}$ we find the same result when varying $\mathfrak{q}$ from 0 to $2\pi/L_x L_y$:
\begin{equation}
\begin{split}
    &\text{Im}\int_0^{2\pi/L_x L_y}d\mathfrak{q}\langle\Psi(\mathfrak{q})|\partial_{\mathfrak{q}}|\Psi(\mathfrak{q})\rangle\\
    &=\text{Im}\sum_{\textbf{p}=1}^{L_xL_y}\int_0^{2\pi/L_x L_y}d\mathfrak{q}\langle\psi(\mathfrak{q})|\partial_{\mathfrak{q}}|\psi(\mathfrak{q})\rangle_{\textbf{p}}=\pi.
\end{split}
\end{equation} Hence we expect this model to have a quantized quadrupole moment $q_{xy}=\frac{e}{2}$ in the limit when the inter-cell interactions dominate. This exactly matches the conclusions of Ref. \onlinecite{ybh2019} for a similar model of hardcore bosons. We also note that in these simple limits, the rank-2 Berry phase boils down to an evaluation of a Berry phases in each plaquette, which seems to have at least a superficial connection to the proposal in Ref. \onlinecite{Araki2019} for Berry phase characterizations of higher order topological insulators. 

Before we move on, let us make comments regarding a realization of a 2D dipole metal in this model. So far our discusions in 2D have been limited to dipole insulator phases. One might imagine that, just as in the 1D ring-exchange model we studied, if we tune $\lambda=t$ so that the inter-cell ring exchange equals the intra-cell ring exchange that the system might be critical and realize a dipole metal point. However, the story here is not clear as when $\lambda$ and $t$ are both non-zero the model is no longer exactly solvable. Indeed there have been works trying to isolate an exciton Bose liquid phase in similar ring exchange models that may have related features\cite{paramekanti2002,motrunich2007,tay2010,tay2011}. We leave a model realization for the 2D dipole metal in this model to future work and move on to discuss an alternative 2D model in which a dipole metal can be realized.

\subsection{2D dipole metal from stacking 1D dipole ladders}

A direct map between the dipole chain Hamiltonian (\ref{eqn:dipole_metal}) and the one-dimensional XY model allows us to construct a 2D model that can be driven through a dipole metal to dipole insulator transition. Consider a system built by stacking dipole chains parallel to $\hat{x}$ into the $y$-direction. Now we introduce a coupling in the $y$-direction that transfers a dipole in its longitudinal direction between rungs of the neighboring dipole ladders (i.e., $y$-pointing dipole moves in the $y$-direction): $d^\dagger_{x,y}d_{x,y+1}$. The overall Hamiltonian now reads:
\begin{equation}
\begin{split}
    H=\sum_{x,y}^{N_x,N_y}&\left(J^x_{\textbf{r}}d^\dagger_{x,y}d_{x+1,y}+J^y_{\textbf{r}}d^\dagger_{x,y}d_{x,y+1}+h.c.\right)\\
    &+U\sum_{x,y}^{N_x,N_y} n_{x,y\uparrow}n_{x,y\downarrow},
    \label{eqn:2d_dipole_metal}
    \end{split}
\end{equation}
where we take $U\gg J^x_{\textbf{r}},J^y_{\textbf{r}}$ to guarantee a well-defined dipole state on every rung of every ladder -- either empty of occupied. Again, this allows us to map dipole creation-annihilation operators to spin-1/2 ladder operators as in Eq. (\ref{eqn:dipoles_to_spin}). In this spin language Eq. \ref{eqn:2d_dipole_metal} takes the familiar form of the 2D XY model on the square lattice:
\begin{equation}
    H=2\sum_{\langle ij\rangle}J_{\langle ij\rangle}\left(S^+_{i}S^-_{j}+S^{-}_iS^+_j\right)
\end{equation}
where the sum runs over every $\langle ij\rangle$ link of the lattice. Now if we double the unit cell in both the $\hat{x}$ and $\hat{y}$ directions by dimerizing the couplings such that $J_{\langle ij\rangle}$ is equal to $t$ for interactions within the expanded unit cell and $\lambda$ for the inter-cell interactions, we arrive at the XY higher order bosonic topological insulator model first considered in Ref. \onlinecite{dubinkin2018higher}. This model can be further mapped\cite{Wang91,dubinkin2018higher} to a free-fermion tight-binding quadrupole model \cite{benalcazar2016} which is in an insulating phase when the inter- and intra- cell couplings $\lambda$ and $t$ are offset and turns is a metal when $\lambda=t$. 

Consider now the coupling of the 2D dipole model Eq. (\ref{eqn:2d_dipole_metal}) to gauge fields. In an external rank-2 gauge field, dipole hopping terms running along the chains pick up a phase $\text{e}^{iA_{yx}}$ as before, while the newly introduced $d^\dagger_{x,y}d_{x,y+1}+h.c.$ terms that tunnel  dipoles between the chains, will pick up a phase factor of $\text{e}^{iA_{yy}}$. Upon mapping this model to the free fermion quadrupole tight-binding model these phase factors become just the ordinary rank-1 phases $\text{e}^{iA_{x}}$ and $\text{e}^{iA_{y}}$ respectively. Importantly, this establishes a direct correspondence between the dipole stiffness Eq. (\ref{eqn:dipole_stiff}) of the dipole model Eq. (\ref{eqn:2d_dipole_metal}) and the usual charge stiffness when calculated for the tight-binding quadrupole model. Whenever the charge stiffness in the tight-binding model is non-vanishing, which is the case exactly when the couplings are non-dimerized, i.e., $\lambda=t$, then we expect a dipole stiffness to take a finite value as well and, therefore, for $\lambda=t$ the model Eq. (\ref{eqn:2d_dipole_metal}) should be a dipole metal. On the other hand, when the couplings are offset, the dipole charge stiffness of the quadrupole model vanishes and the dipole stiffness must vanish as well signifying that the ground state of the dipole Hamiltonian is a \emph{dipole-insulating} state.

\section{Dipole Incompressibility}
We have provided some formal definitions for dipole metals and insulators (and even one example of a dipole superconductor), but we have not provided as much physical intuition behind these definitions, except in analogy to charge metals and insulators. Another way to describe dipole insulators, that may provide additional physical context, is in terms of \emph{dipole incompressibility}. We can make the following simple observation: since dipole moments couple to the gradient of the electromagnetic field, we expect that $\partial_i A_0$ can shift the chemical potential for dipoles similar to how $A_0$ shifts the chemical potential for electrons. So, for the $\hat{x}$-dipoles, we have $\mu_{d^x}\equiv \partial_x A_0\propto E_x$. Therefore, a dipole incompressibility condition reads (for $\hat{x}$-dipoles having polarization $P_x$):
\begin{equation}
    \frac{d P_x}{d\mu_{d^x}}\propto \frac{d P_x}{d E_x}=\chi_e,
    \label{eqn:dip_incompressibility}
\end{equation}
where $\chi_e$ is the electric polarizability. Therefore, a dipole incompressibility condition simply translates to a simple equation: $\chi_e=0$. Thus, intuitively, a dipole insulator is a charge insulator (dielectric) that does not polarize when a small electric field is applied.

We can put this condition to test in our 2D dipole insulator model. As we can easily check, the ground state given by (\ref{eqn:dipole_p_gs}) and (\ref{eqn:overall_gs}) has zero polarization in either the $\hat{x}$ or $\hat{y}$-directions. We can apply a constant external electric field $E_x$ by introducing a constant gradient of $A_0$: $A_0(\textbf{r})=E_0 x$. What we immediately find is that the ground state on each individual plaquette is also an eigenstate of the electric field operator. Thus, we expect the ground state \emph{wavefunction} will not get any corrections and therefore, the polarization will not change at all. As such, we find the electric susceptibility is indeed exactly zero in this 2D model. 

It is worth examining the structure of the ground state of this model a bit more closely to find a better understanding why $\chi_e=0$. The reason turns out to be simple. From the point of view of the lattice, turning on the electric field $E_x$ is just changing the potential energy at each site such that the potential on a site with coordinates $(x,y)$ is equal to $E_0 x$. So, for a two-fermion state of electrons with coordinates $(x_1,y_1)$ and $(x_2,y_2)$ the operator $\hat{V}=E_x$ simply changes the potential energy of such state by $\Delta \epsilon=E_0 x_1+E_0 x_2$. However, as we discussed above, the ground state on each plaquette is a linear combination of two states that have electrons sitting at the opposite corners of a plaquette, therefore, for a plaquette attached to a unit cell with coordinates $(x_0,y_0)$ we have:
\begin{equation}
    \hat{V}\ket{\psi}_p=E_0(x_0+(x_0+1))\ket{\psi}_p
    \label{eqn:perturb_ex}
\end{equation}
And we find the same result for $E_y$.
By this measure, we conclude that any state represented by a linear combination of states with the same average coordinate, is an eigenstate of the electric field operator. So, if we perturb our ground state by an operator that conserves this average coordinate, for example by turning on a small amount of non-zero intra-cell coupling $t$, we will find that the perturbed state is also an eigenstate of $E_x$ and, therefore, it also has vanishing electric susceptibility. Thus, by this measure, our model is a dipole insulator, even away from a zero correlation length limit.

\section{Discussion and Conclusion}
Our results represent a refinement of the notion of an insulating system: a well-defined $n$-th order multipole insulator can be further identified as either an $(n+1)$-th order multipole metal or insulator. In this work we focused on dipole metals and insulators, but these concepts can be extended, essentially \emph{mutatis mutandis}, to higher order multipoles. In each case we can imagine a hierarchy of possible localization-delocalization transitions where, say, the $m$-th order multipole becomes delocalized, hence destroying any well-defined notion of $n$-th order multipole moments for $n>m.$ However, while concepts like the $n$-th order multipole stiffness and the Berry phase associated with the $n$-th multipole twist operator $U_{X_1 X_2...X_n}$ can be introduced, there are no readily available models to test them (other than perhaps a 3D octupole model with subsystem symmetry in Ref. \onlinecite{ybh2019}). Thus, this leaves many open questions, especially when trying to identify models that represent $n$-th order multipole metals.

To find our results we have re-purposed several key concepts used to distinguish between regular charge metals and insulators to study the analogous dipole conserving systems. Specifically, we defined a dipole stiffness $D_d$ and dipole localization area $\lambda_d$ - natural extensions of charge stiffness (or Drude weight), and charge localization length. We also discussed a dipole incompressibility condition (the electric polarizability vanishes) that parallels the incompressibility condition in ordinary charge insulators.
We also proposed that the expectation value of the the twist operator $z_{XY}=\langle U_{XY}\rangle$ acts as a universal criterion for the distinction between dipole metals where $|z_{XY}|\to 0,$ and dipole insulators where $|z_{XY}|\to 1$ in the thermodynamic limit. Additionally, we showed that the Berry phase, related to the twist introduced by $U_{XY}$ (or alternatively a shift of the rank-2 gauge field), can be used to distinct between the gapped phases with different quadrupolar polarizations. 

In addition to the conceptual developments, we provided several models to test the concepts. Using ring-exchange terms as natural building blocks for charge-insulating systems having dipole conservation, we constructed a handful of toy-models exhibiting dipole metallic or dipole insulating phases. For our one-dimensional model we discussed its realization in a cold-atom context where it could be possible to experimentally probe simple dipole metal and insulator phases. Having noticed that ring-exchange terms naturally couple to a rank-2 gauge field we also applied a continuum description of dipole phases and provided an example of a system that, while having a vanishing charge Drude weight, has a non-zero dipole stiffness $D_d$.

These results open a considerable amount of questions for further study. One important avenue for exploration is finding potentially exotic phases of matter associated with multipolar insulators. Specifically, is there a rich landscape of phases similar to those discovered in charge insulators over the past few decades? Additionally, there is no exact answer to the question of the role of fractonic phases. That is,  must fracton theories inevitably emerge when one studies multipole insulators, or are they just a subclass of multipole insulating phases? Next, there are many open questions regarding the connection to higher order multipole topological insulators. Recent work has shown a connection between some classes of subsystem protected (fracton) topological phases, and higher order topological insulators when the subsystem symmetry is broken down to a global symmetry\cite{ybh2019}. Thus, the higher order topological insulators do not have exact dipole conservation at the Hamiltonian level, but there may be some notion of an emergent dipole conservation. Finally, there are questions regarding connections between the dipole insulator and Bose metal phases such as the exciton Bose liquid. We leave these questions for future works.

\begin{acknowledgements}
We thank F. J. Burnell, L. K. Wagner, W. A. Wheeler, Y. You for discussions on related projects. TLH also thanks F. Pollmann for a useful discussion. OD and TLH
thank the US National Science Foundation under grant
DMR 1351895-CAR, and the MRSEC program under NSF Award Number DMR-1720633 (SuperSEED) for support  JM is supported by the National Science Foundation Graduate Research Fellowship Program under Grant No. DGE – 1746047. TLH also thanks the National Science
Foundation under Grant No.NSF PHY-1748958(KITP)
for partial support at the end stage of this work during the Topological Quantum Matter program.

\end{acknowledgements}
\appendix
\section{Rank-1 vs. rank-2 background field couplings}\label{app:rank1rank2}
In this Appendix we will discuss the similarities and differences between coupling dipoles to the derivative of a background rank-1 field $\partial_i A_j$ and coupling dipoles to a background rank-2 field $A_{ij}$. 

First, we will show how a 2 particle system can naturally couple to $\partial_i A_j$. We will do this by considering a many body Schrodinger equation for an electron and hole wavefunction $\Psi(\bm{x}_e, \bm{x}_h) $
\beq
\nonumber H = \frac{1}{2m}\left|\bm{p}_e-e\bm{A}(\bm{x}_e)\right|^2 &+& \frac{1}{2m}\left|\bm{p}_h+e\bm{A}(\bm{x})\right|^2\\ &+& V(\bm{x}_e,\bm{x}_h),
\eeq
where $V$ is an interaction term between the electron and hole, and we have set the electron and hole masses equal to each other. 

It will be useful to to rewrite the Hamiltonian in terms of the center of mass ($ \bm{x}_{cm}  = (\bm{x}_e+\bm{x}_h)/2 $) and relative coordinate as ($ \bm{x}_{r}  = \bm{x}_e-\bm{x}_h $) as
\beq
\nonumber H &=& \frac{1}{m}\left|\bm{p}_{cm}-e\bm{A}(\bm{x}_{cm}+\frac{\bm{x}_r}{2})+e\bm{A}(\bm{x}_{cm}-\frac{\bm{x}_r}{2})\right|^2\\\nonumber &\phantom{=}& + \frac{1}{2m}\left|\bm{p}_r-e\bm{A}(\bm{x}_{cm}+\frac{\bm{x}_r}{2})-e\bm{A}(\bm{x}_{cm}-\frac{\bm{x}_r}{2})\right|^2\\ &\phantom{=}& + V(\bm{x}_e,\bm{x}_h).
\eeq
We will now assume that the interaction $V$ results in the electron and hole forming a bound state with a definite dipole moment $\bm{d}$. This is accomplished by $V = V_0 |\bm{x}_r - \bm{d}|^2$. In the limit where $V_0\rightarrow \infty$, $\Psi$ will satisfy 
\beq
|\Psi(\bm{x}_e, \bm{x}_h)|^2 \propto \delta(\bm{x}_e - \bm{x}_h - \bm{d}),
\label{eq:DipoleRel}
\eeq
and we can subsequently set $\bm{x}_r = \bm{d}$ and ignore the fluctuations in the relative coordinate. 
After Taylor expanding the vector potential $\bm{A}$, the center of mass part of the Hamiltonian becomes
\beq
H &=& \frac{1}{m}\left|\bm{p}_{cm}-e d_i \partial_i  \bm{A}(\bm{x}_{cm})\right|^2.
\eeq
This expansion is valid provided that the derivatives of the background field $\bm{A}$ do not strongly fluctuate on length scales of order the dipole length $|\bm{d}|$. From this we can conclude that free dipoles can couple to the derivative of the rank-1 gauge field: $\partial_i A_j$. 

In the study of fractonic phases of matter, it has also been shown that dipole excitations naturally couple to symmetric rank 2 gauge fields $A_{ij}$. The rank 2 gauge fields transform as $A_{ij} \rightarrow A_{ij} + \partial_i \partial_j \gamma$ (note that this is the same gauge transformation as $\partial_i A_j$). When considering dipole dynamics, we thereby have the choice of either considering a system where the dipoles couple to the derivative of a background rank-1 gauge field $\partial_i A_j$ or a background rank-2 gauge field $A_{ij}$. We will consider this choice in two different contexts. First, on systems with open boundaries and second on systems with periodic boundaries.

For systems with open boundaries, there is no distinction between coupling a dipole to the derivative of a background rank-1 gauge field and a background rank-2 gauge field. This is because we can always equate the rank-2 gauge field with the symmetric derivative of the rank-1 gauge field, i.e., $A_{ij} = (\partial_i A_j + \partial_j A_i)/2 $. In reverse, we can also equate the integral of the rank-2 gauge field with the rank-1 gauge field $A_j =  \int dx_i A_{ij}$. %These two equations are consistent, since $A_{ij} = (\partial_i \int dx_i A_{ij} + \partial_j \int dx_j A_{ji})/2 = (A_{ij} + A_{ji})/2 = A_{ij}$, where we have used the the fact that $A_{ij}$ is symmetric. 

On periodic geometries, however, these descriptions are seemingly not equivalent. When considering the derivative of a background rank-1 gauge field on a periodic system, we require that the background rank-1 gauge field $A_j$ is periodic, while for the rank-2 description, we require that the rank-2 gauge field $A_{ij}$ is periodic. In particular, for a given background periodic rank-2 gauge field $A_{ij}$, there might not be a periodic rank-1 gauge field $A_{i}$ such that $A_{ij} = (\partial_i A_j + \partial_j A_i)/2 $. In other words, $\int dx_i A_{ij}$ may not satisfy periodic boundary conditions. 

As a specific example of this, let us consider the rank-2 gauge field $A_{xy}(x,y) = \alpha$(const.). On open boundaries, one can define the rank-1 gauge field $A_x(x,y) = \alpha y$, $A_y(x,y) = \alpha x$, and clearly $A_{xy} = (\partial_x A_y + \partial_y A_x)/2 $. On periodic boundary conditions, $A_{xy}(x,y) = \alpha$ is still a viable background field configuration. However, in general, $A_x(x,y) = \alpha y$, $A_y(x,y) = \alpha x$ is not a viable background field configuration configurations for periodic boundary conditions, since it is not invariant under $x \rightarrow x+L_x$ and $y \rightarrow y+L_y$. 

The difference between rank 1 and rank 2 couplings is thereby meaningful if we consider periodic boundaries. For open boundaries, on the other hand, it is possible to equate the two. We note that this is only true when considering background fields. If we instead consider the rank-1 and rank-2 fields to be dynamic, the situation is entirely different, since the two types of fields have different path integrals and quantum theories.   

\section{Dipole stiffness}\label{app:linearresponse}
Let us derive the relationship between the dipole stiffness and dipole conductivity density. First, we couple our system to a rank-2 gauge field $A_{ij}$ via 
\begin{equation}
	H=H_0+\Delta H,\quad \Delta H=-A_{ij} J^d_{ij},
\end{equation} 
where $J^d$ is the dipole current. Note that $J^d$ may depend on $A_{ij}$. In general, we expect $J^d_{ij} = J^0_{ij} - A_{ij} J^A_{ij}/2$, where both $J^0_{ij}$ and $J^A_{ij}$ are independent of $A_{ij}$. Using the Kubo formula, the response of the dipole current density $ j^d_{ij} \equiv J^d_{ij}/V$ to this pertubation is
\beq
&&\langle j^d_{ij}(t)\rangle \equiv \frac{\langle J^d_{ij}(t)\rangle}{V}= (\rho_d(t) + \chi_d(t))A_{ij}(t)
\eeq
where $\chi_d$ is the retarded Green function of the dipole current-dipole current correlation function:
\begin{equation}
	\begin{split}
		%&\langle J^d_{xy}(t)\rangle
		%-\frac{1}{i \omega}\int_{-\infty}^t dt'\langle 0|[J^d_{xy}(t'),J^d_{xy}(t)]|0\rangle k e^{-i\omega t'}\\
		%\chi_d=\frac{1}{\omega}\left(\int_{0}^\infty d\tau\langle 0|[J^d_{xy}(0),J^d_{xy}(\tau)]|0\rangle e^{i\omega \tau}\right) E_{xy} e^{-i\omega t}
		\chi_d (t)= - \frac{i}{\hbar V} \int_{-\infty}^t dt' \langle 0|[J^d_{xy}(t),J^d_{xy}(t')]|0\rangle_0,
	\end{split}
	\label{}
\end{equation}
where the subscript $0$ indicates that the correlation function is calculated with $A_{ij} = 0$. The quantity $\rho_d(t) = \langle J^A_{ij} (t)\rangle_0/V$ can be interpreted dipole analog of the current density. %For the Hamiltoian @@@, $\rho_d = |\Phi|^4$.

Let us now restrict ourselves to the case $(i,j) = (x,y)$. Other values of $i$ and $j$ are determined in an analogous way. To derive a dipole conductivity, we will introduce the rank 2 electric field $E_{xy}$ that is the canonical conjugate of $A_{xy}$:
\begin{equation}
	E_{xy} = -\frac{\partial}{\partial t} A_{xy}.
	\label{}
\end{equation}
In terms of frequency, $E_{xy}(\bm{x},\omega) = i \omega A_{xy}(\bm{x},\omega)$, and:
\beq
\nonumber \langle j^d_{xy}(\omega)\rangle &=& \left(\frac{\rho_d(\omega)}{i\omega} + \frac{\chi_d(\omega)}{i\omega}\right)E_{xy}\\ &=& \sigma_d(\omega) E_{xy}
\eeq
%In actual calculations, we will be interested in the spatially averaged dipole momentum, 
%\beq
%\sigma_d(\omega) = \frac{1}{V}\int d^3 \bm{x} %\sigma_d(\bm{x},\omega),
%\eeq
%where $V$ is the system volume.
where $\sigma_d$ is the dipole conductivity. In general this quantity will be a tensor, but since we are only considering $(i,j)= (x,y)$, the tensor structure is irrelevant to our present derivation.

The Fourier transform of the position-averaged correlator $\chi_d(\omega)$ can be evaluated by inserting a basis of energy states between dipole current operators:
\begin{equation}
	\begin{split}
		\chi_d(\omega)=-\frac{i}{V\hbar} \int_{0}^\infty d\tau &e^{i\omega \tau} \sum_n|\langle 0|J^d_{xy}|n\rangle_0|^2\times\\
		&\times\left(e^{i(E_n-E_0)\frac{t}{\hbar}} - e^{-i(E_n-E_0)\frac{t}{\hbar}}\right)
	\end{split}
	\label{}
\end{equation}
where $\sum_n|\langle 0|J^d_{xy}|n\rangle_0|^2$ is again evaluated with $A_{xy} = 0.$ This gives:
\begin{equation}
	\begin{split}
		\sigma_d(\omega)=\frac{\rho_d}{i \omega} -\frac{1}{i V \omega}\sum_{n\neq 0}|\langle 0|J^d_{xy}|n\rangle_0|^2&\left(\frac{1}{\hbar\omega+i\varepsilon+E_n-E_0}\right.\\
		&\left.-\frac{1}{\hbar\omega+i\varepsilon+E_0-E_n}\right).
	\end{split}
	\label{}
\end{equation}
Due to the singularities at $\pm \hbar\omega = E_n-E_0$, we can express the real part of the conductivity as:
\begin{equation}
\begin{split}
	\text{Re}\sigma_d(\omega)=\frac{\pi}{V \omega}\sum_{n\neq 0}|\langle 0|J^d_{xy}|n\rangle_0|^2[\delta(\hbar\omega-E_n + E_0)\\-\delta(\hbar\omega+E_n - E_0)].
	\label{}
\end{split}
\end{equation}
 On the other hand, for the imaginary part we have:
\begin{equation}
	\lim_{\omega\to 0}\text{Im}\sigma_d(\omega)= -\frac{\rho_d}{\omega}+\frac{2 }{V \omega}\sum_{n\neq 0}\frac{|\langle 0|J^d_{xy}|n\rangle_0|^2}{E_n-E_0}.
	\label{}
\end{equation}
This allows us to define the corresponding dipole stiffness as:
\begin{equation}
	D_d=\pi \lim_{\omega\to 0}\text{Im}\omega\sigma_d(\omega).	
	\label{}
\end{equation}
Which gives us:
\begin{equation}
	D_d=-\pi\rho_d +\frac{2\pi}{V}\sum_{n\neq 0}\frac{|\langle 0|J^d_{xy}|n\rangle_0|^2}{E_n-E_0}.
	\label{eqn:drude_weight_expression}
\end{equation}
On the other hand, we can couple our system to a constant rank-2 background field $A_{xy} \equiv \mathfrak{q}$ via
\begin{equation}
	\Delta H= -\mathfrak{q} J^d_{xy} =  -\mathfrak{q} J^0_{xy}+\frac{1}{2}\mathfrak{q}^2 J^A_{xy},
	\label{}
\end{equation}
and then calculate the second order correction to the energy. Using
\begin{equation}
	E^{(1)}=-\langle 0| J^d_{xy} |0\rangle, 
\end{equation}
and
\begin{equation}
	\begin{split}
	E^{(2)}&= \frac{1}{2} \mathfrak{q}^2 \langle 0 | J^A_{xy} | 0 \rangle_0 - \mathfrak{q}^2\sum_{n\neq 0}\frac{|\langle 0|J^0_{xy}|n\rangle_0|^2}{E_n-E_0},
	\end{split}
	\label{}
\end{equation}
where $V$ is the volume of the sample,  we can show that:
\begin{equation}
	\left.\frac{\partial^2 E}{\partial \mathfrak{q}^2}\right|_{\mathfrak{q}=0}=\langle 0 | J^A_{xy} | 0 \rangle_0-2\sum_{n\neq 0}\frac{|\langle 0|J^D_{xy}|n\rangle_0|^2}{E_n-E_0}.
	\label{}
\end{equation}
Combining this expression with (\ref{eqn:drude_weight_expression}) we finally get:
\begin{equation}
	D_d=-\frac{\pi}{V}\left.\frac{\partial^2 E}{\partial \mathfrak{q}^2}\right|_{\mathfrak{q}=0}.
	\label{eqn:ddw1_expression}
\end{equation}

\section{Dipole equations of motion}\label{app:dipolewave}
Let us start with a simplified version of Pretko's Lagrangian
\begin{equation}
    \mathcal{L}=|\partial_t \Phi|^2-m^2|\Phi|^2-\lambda|\Phi\partial_x\partial_y\Phi-\partial_x\Phi\partial_y\Phi|^2
    \label{eqn:dipole_lagrangian}
\end{equation}
where we kept only the off-diagonal - $xy$ - kinetic terms that govern transversal movement of dipoles. The Euler-Lagrange equations when $\mathcal{L}$ includes higher-order derivatives read:
\begin{equation}
    \frac{\delta\mathcal{L}}{\delta \Phi}-\partial_\mu \frac{\delta\mathcal{L}}{\delta (\partial_\mu\Phi)}+\partial_\mu\partial_\nu \frac{\delta\mathcal{L}}{\delta (\partial_\mu\partial_\nu\Phi)}=0.
\end{equation}
For the Lagrangian (\ref{eqn:dipole_lagrangian}) this equation obtained via variation with respect to $\Phi^\dagger$ read:
\begin{equation}
    \begin{split}
        &-2\partial_t^2\Phi-2m ^2\Phi-4\lambda(\partial_x\partial_y\Phi^\dagger)D_{xy}[\Phi]-2\lambda(\partial_y\Phi^\dagger)\partial_x D_{xy}[\Phi]\\
        &-2\lambda(\partial_x\Phi^\dagger)\partial_y D_{xy}[\Phi]-\lambda\Phi^\dagger\partial_x\partial_y D_{xy}[\Phi]=0
    \end{split}
\end{equation}
where we used a shorthand notation for the rank-2 covariant derivative $D_{xy}[\Phi]=\Phi\partial_x\partial_y\Phi-\partial_x\Phi\partial_y\Phi$. Although we are not going to task ourselves with solving this equation, we can easily check that a set of non-trivial solutions is given by:
\begin{equation}
    \Phi_{\mathfrak{q}}(t,x,y)=\varepsilon\text{e}^{i\left(\mathfrak{q}xy-\omega t\right)}
\end{equation}
where $\varepsilon$ is the dimensionful amplitude of the field. This leads us to write the following dispersion relation for these 'dipole-wave' %{\bf{TLH: Why is it called a dipole wave? How is the dispersion relation calculated? OSD: We can change this terminology. The function $\Phi_{\mathfrak{q}}$ can be multiplied by a phase factor $\text{e}^{i\alpha(x)}$ and it's going to stay the solution of the EOMs with the same dispersion relation. This additional symmetry implies dipole conservation and hence we constructed the closest analogue of a plane wave solution for a dipole-conserving system hence the name. The dispersion relation is obtained by inserting $\Phi_{\mathfrak{q}}$ into the EOMs and solving them.}}
solutions:
\begin{equation}
    \omega^2-m^2-\lambda|\varepsilon|^2\mathfrak{q}^2=0.
\end{equation}
Note that with addition to regular global phase rotations, phase rotations that depend only on one coordinate leave the Lagrangian invariant as well as don't move the solution to the equation of motion off-shell. In other words, we have a $U(1)\times U(1)$ subsystem symmetry given by:
\begin{equation}
    \Phi\to\text{e}^{i\alpha(x)}\Phi,\ \Phi\to\text{e}^{i\beta(y)}\Phi
\end{equation}
where $\alpha(x)$ and $\beta(y)$ are arbitrary functions of $x$ and $y$ respectively.

\section{Dipole correlation area for 2D dipole insulator with periodic boundary conditions}\label{app:zmagnitude}
Let us compute the value of $|z_{XY}|=|\langle U_{XY}\rangle|$ for the ground state of the 2d dipole insulator on a lattice with periodic boundary conditions. We find that the terms coming from the boundary contribute quite differently from the bulk. The full expression for the $|z_{XY}|$ reads:
\begin{equation}
    \begin{split}
    |z_{XY}|&=\left|\cos\left(\frac{\pi}{N_x N_y}\right)\right|^{(N_x-1)(N_y-1)}\\
    &\times\left| \cos\left(\frac{\pi(N_x-1)}{N_xN_y}\right) \right|^{N_y-1}\times\left| \cos\left(\frac{\pi(N_y-1)}{N_xN_y}\right)\right|^{Nx-1}\\
    &\times\left|\cos\left(\frac{\pi(N_x-1)(N_y-1)}{N_x N_y}\right)\right|.
    \end{split}
    \label{eqn:expectation_zxy}
\end{equation}
If we compute the dipole correlation area from this expression we find that it is not as well-behaved, i.e., we find
\begin{equation}
\begin{split}
    \lambda^2_d&=-\frac{N_d}{4\pi^2\rho_d^2}\log\left|z_{XY}\right|^2\\
    &\approx \frac{a^4}{4}N_x N_y\left[-\frac{1}{N_xN_y}+\frac{N_x^2+N_y^2}{N^2_xN^2_y}+\frac{1}{N_x}+\frac{1}{N_y}\right]\\
    &\approx \frac{a^4}{4}(N_x+N_y)+ O(1),
    \end{split}
\end{equation}
which leads to  $\lambda_d=\frac{a^2}{2}\sqrt{N_x+N_y}$. This quantity diverges with the lattice size, and is not the correct physical result. Viewing this as an artifact of the non-periodic behaviour of $U_{XY}$ at the boundary, we could evaluate Eq. \ref{eqn:expectation_zxy} using open boundary conditions. This is the result already presented in the main text. 

Alternatively, we propose the following way to resolve this issue. Consider the 2D dipole insulating Hamiltonian on an $L\times L=Na\times Na$ square lattice with periodic boundary conditions. Now imagine we treat every vertical column of the lattice as a large supercell. This formal redefinition changes the dipole filling number: $\rho_d=N/a^2$ and requires us to use the following operator instead of $U_{XY}$ to calculate $z_{XY}$:
\begin{equation}
    \tilde{U}_{XY}=\exp\left[\frac{2\pi i}{aL}\sum_{i,\alpha} x_{i,\alpha}y_{i,\alpha}\hat{n}_{i,\alpha}\right],
\end{equation}
where the index $i$ runs over all the unit cells, and $\alpha$ runs over the sites within a unit supercell. Note that this new operator is completely periodic and, additionally, it can be expressed as $\tilde{U}_{XY}=(U_{XY})^N$. Now, computing $|z_{XY}|$ for the ground state of the 2D dipole insulator in the limit when $\lambda=1$, $t=0$ we find:
\begin{equation}
    |\tilde{z}_{XY}|=\left|\cos\left(\frac{\pi}{N}\right)\right|^{N\times N},
\end{equation}
and the dipole correlation area is given by:
\begin{equation}
    \lambda^2_d=-\frac{N_d}{4\pi^2\rho_d^2}\log\left|z_{XY}\right|^2\approx \frac{a^4 N^2}{4\pi^2 N^2}\pi^2=\frac{a^4}{4},
\end{equation}
which gives: $\lambda_d=a^2/2$. This is the same value of the dipole correlation area that we obtained when we evaluated the magnitude using open boundary conditions and the original $U_{XY}$ operator.

\bibliography{References}

\end{document}